\documentclass[twocolumn]{jpsj3}
\usepackage{ulem}
\usepackage{color}
\usepackage{braket}
\usepackage{cite}
\usepackage{amsmath}
\definecolor{mygreen}{rgb}{0.0,0.7,0.0}
\newcommand{\txr}[1]{\textcolor{black}{#1}}
\newcommand{\red}[1]{\textcolor{black}{#1}}

\title{ Excitons and \red{Dark} Fermions as Origins of Mott Gap, Pseudogap and Superconductivity in Cuprate Superconductors \\ 
--- General \red{Concept} and Basic Formalism Based on Gap Physics}

\author{Masatoshi Imada and Takafumi J. Suzuki}
\inst{%
 Department of Applied Physics, University of Tokyo, 7-3-1 Hongo, Bunkyo-ku, Tokyo 113-8656
}%

\abst{Theory of doped Mott insulators is revisited in the light of recent understanding on the singular self-energy structure of the single-particle Green's function. The unique pole structure in the self-energy induces the high-temperature superconductivity in the anomalous part, while it generates Mott gap and pseudogap in the normal part. Here, we elucidate that fractionalization of electrons, which is exactly hold in the Mott insulator in the atomic limit, more generally produces the emergent Mott-gap fermion and dark (hidden) fermions.  It does not require any spontaneous symmetry breaking. The two gaps are the consequences of the hybridization of these two fermions with quasiparticles. We further propose that the Mott-gap \red{fermion} and dark fermions are the fermionic component of Frenkel- and Wannier-type excitons, respectively, which coexist in the doped Mott insulator.  
The Bose-Einstein condensation of the Frenkel-type excitons allowed without spontaneous symmetry breaking holds a key for understanding the unique pole structure and the pseudogap through the instantaneous hybridization between the \red{fractionalized} quasiparticle and the \red{dark} fermion \red{in analogy with the Mott gap}. We argue that the high-$T_{\rm c}$ superconductivity is ascribed to the dipole attraction of the Wannier-type excitons.
The gap formation mechanism is compared with that caused by conventional spontaneous symmetry breaking known over condensed matter and elementary particle physics including quantum chromodynamics.  
We propose a theoretical framework and discuss experimental tests to analyze this idea and concept.
}

\begin{document}
\maketitle

%
\section{Introduction}
%
Understanding physics and the mechanism of superconductivity that may allow designing higher superconducting critical temperature $T_{\rm c}$ achieved in cuprate superconductors are still open issues and progress has been continuing since its discovery. The origin of the superconducting phase itself
, as well as that of the pseudogap found above $T_{\rm c}$ in the normal phase~\cite{Yasuoka,Alloul,RMP} in the underdoped region near the Mott insulator of mother compounds have not reached complete understanding. 
In addition to the $d$-wave superconductivity, increasing experimental indications show various competing orders including spatial electronic inhomogeneity such as stripe-like or nematic charge order~\cite{Tranquada1995,Tranquada1997,Yamada1998,Fink2011,Ghiringhelli,Comin,Peng,SatoNematic} and electronic mesoscopic phase separation in this underdoped region\cite{davis}.  
In this article, we propose a novel and consistent  mechanism on this issue and a formalism that allows testing the proposal numerically to gain insight into future possible experimental verifications.  

To understand the origin of the pseudogap in the cuprates,  momentum $k$- and frequency $\omega$-dependent single-particle electronic Green's function  
 \begin{eqnarray}
G(k,\omega)=\frac{1}{\omega-\epsilon(k)-\Sigma (k,\omega)},
\label{Gkw} 
\end{eqnarray}
has been studied with the bare band dispersion $\epsilon (k)$, and self-energy $\Sigma (k,\omega)$\cite{Maier,Senechal,Civelli2005,Berthod,Stanescu,Konik,YRZ,HauleKotliar,Norman,Sakai2009,Civelli2009PRL,Civelli2009,Sakai2010,Gull,Sakai2016,Sakai2016_2} for the Hubbard model defined by
\begin{eqnarray}
H_{\rm Hub} &=&-\frac{1}{2}\sum_{i,\delta,\sigma}t_{\delta}[c_{i,\sigma}^{\dagger}c_{i+\delta,\sigma}+{\rm H.c}] -\mu \sum_{i,\sigma}n_{i,\sigma} \nonumber \\
&&+U\sum_{i}c_{i,\uparrow}^{\dagger}c_{i,\uparrow}c_{i,\downarrow}^{\dagger}c_{i,\downarrow} ,
\label{Hubbard} 
\end{eqnarray}
on the square lattice.  
Here, $c_{i,\sigma}^{\dagger} (c_{i,\sigma})$ creates (annihilates) an electron on the site  $i$ with spin $\sigma$ and $n_{i,\sigma}$ is the number operator. 

In particular, singular structure of normal and anomalous self-energy, represented by poles have been extensively studied\cite{Stanescu,Konik,YRZ,Sakai2009,Civelli2009,Sakai2010,Gull,Sakai2016,Sakai2016_2}.
The structure of 
the self-energy $\Sigma(k,\omega)$ is crucially important in understanding the pseudogap and the superconductivity.
It is also important in understanding physics emerging in the anomalous metals around the Mott insulator.  

The single-particle gap generated by interaction effects is in general represented by poles of $\Sigma$ near $\omega=0$ (Fermi level), because the divergence of $\Sigma$ at a pole leads to a zero of the Green's function in Eq.(\ref{Gkw}) and leads to suppression of the density of states ${\rm Im} G(\omega=0)$ at the Fermi level.
The divergence of the self-energy by itself signals the breakdown of the standard perturbation theory in terms of the interaction.

Aside from the trivial single-particle gap formed under the periodic potential of nuclei in crystals, 
nontrivial excitation gap arising from the self-energy structure is found in many condensed matter systems when a symmetry is spontaneously broken as in the cases of charge order, antiferromagnetic order and superconducting states.    

Let us consider very general mechanisms of the gap or mass generation originating from many-body physics.
In the magnetic or charge orders, the mean field decoupling of the Coulomb repulsion generates the excitation gap.
The formation of a nontrivial single-particle excitation gap in interacting fermion systems can be interpreted by the emergent ``hybridization gap" in the effective single particle problem given from the Hamiltonian
\begin{eqnarray}
H&=&\sum_{k,\sigma,\sigma'}[ \epsilon_c (k)c_{k,\sigma}^{\dagger}c_{k,\sigma} +\Lambda (k) (c_{k,\sigma}^{\dagger}d_{k,\sigma} +{\rm H.c.}) \nonumber \\
&+&\epsilon_d (k)d_{k,\sigma}^{\dagger}d_{k,\sigma} ] .
\label{TCfermionNormal} 
\end{eqnarray}
 Here, the fermion represented by $d$ with the dispersion $\epsilon_d(k)$ is hybridizing with the fermion of our interest $c$ with the dispersion $\epsilon_c(k)$ at the momentum $k$ in a form of a noninteracting Hamiltonian. 
 The direct hybridization gap at each momentum is given by
\begin{eqnarray}
\Delta_{HG}=\sqrt{(\epsilon_c(k)-\epsilon_d(k))^2+4\Lambda (k)^2}.
\label{HGap} 
\end{eqnarray} 

The solution of Eq.(\ref{TCfermionNormal}) gives the Green's function for $c$ in the form of Eq.(\ref{Gkw}) with
\begin{eqnarray}
\Sigma(k,\omega)=\frac{\Lambda (k)^2}{\omega-\epsilon_d(k)}.
\label{TCMGkw} 
\end{eqnarray}
Equation (\ref{TCMGkw}) indicates that the pole of the self-energy emerges at $\omega= \epsilon_d(k)$, namely at the bare dispersion of the fermion $d$. This pole generates the zero of $G$ and a gap in the density of states of the fermion $c$ known as the hybridization gap.

The form (\ref{TCfermionNormal}) 
emerges by the mean-field decoupling of interacting fermions such as the Hubbard model (\ref{Hubbard}). The Coulomb interaction between $c$ and $d$ represented by
$g c_{k,\sigma}^{\dagger}d_{k,\sigma} c_{q,\sigma'}^{\dagger}d_{q,\sigma'} $ is decoupled to 
$\Delta c_{k,\sigma}^{\dagger}d_{k,\sigma}$ with $\Delta= g\langle c_{q,\sigma'}^{\dagger}d_{q,\sigma'}\rangle $ if the spontaneous symmetry breaking takes place with the order parameter 
$\langle c_{k,\sigma}^{\dagger}d_{k,\sigma}\rangle $.
 In the charge order and antiferromagnetic order, $d_{k,\sigma}$ hybridizing with $c_{k,\sigma}$ is nothing but 
the fermion $c$ itself at different wave number $d_{k,\sigma}=c_{k+Q,\sigma'}$, where $Q$ is the ordering wave vector and $\sigma'$ can be the same spin as $\sigma$ for the charge order and antiferromagnetic order, while it can be $\sigma'=-\sigma$ for the antiferromagnetic order aligned perpendicular to the spin quantization axis $z$.  In the case of the superconductivity, $d_{k,\sigma}=c^{\dagger}_{-k,\sigma'}$ forms a Cooper pair
$c_{k\sigma}c_{-k-\sigma}$ in Eq.(\ref{TCfermionNormal}).

The same mechanism applies in the quantum chromodynamics (QCD)~\cite{Weinberg}, where $c$ is a quark and $d$ is an antiquark operator. For the strong interaction case, nearly the SU(3)-symmetric  representation of up, down and strange quarks constitutes the QCD Lagrangian, and after tracing out the gluon, the resultant quark interaction term is decoupled by the  quark-antiquark condensation through the chiral symmetry breaking and an emergent hybridization of quark and antiquark appears. Again the hybridization gap generates the mass of quarks through the Nambu-Jona Lassinio mechanism\cite{Nambu-Jona_Lassinio}.  

Another mechanism of the mass (gap) generation is found in fermion-boson coupled systems essentially represented by
\begin{multline}
H=\sum_{k}[ \epsilon_c (k)c_{k}^{\dagger}c_{k} 
 \\ 
+ \sum_{q}\Lambda (k,q) (c_{k}^{\dagger}d_{k+q}(b_q+b^{\dagger}_{-q}) +{\rm H.c.}) +\epsilon_d (k)d_{k}^{\dagger}d_{k} ] 
\label{TCfermionboson} 
\end{multline}
If the bosons $b$ condense by the Bose-Einstein condensation, 
$\langle b \rangle =\langle b^{\dagger} \rangle^*\ne 0$, we again reach the form of Eq.(\ref{TCfermionNormal}).
This is the case of the weak interaction, where the boson is either W boson or Z boson and $c$ and $d$ are hadrons such as nucleons. In the case of the strong interaction, the gluon condensation may also generate a gap originated from the same mechanism. 

All of these gap generation require spontaneous symmetry breaking.
On the other hand,
in the atomic limit of the Hubbard model (\ref{Hubbard}), $t_{\delta}\rightarrow 0$, at half filling $\mu=U/2$, the Green's function is exactly given by~\cite{Sakai2016_2} 
 \begin{eqnarray}
G(k,\omega)=\frac{1}{2}\left[\frac{1}{\omega+\frac{U}{2}}+\frac{1}{\omega-\frac{U}{2}}\right],
\label{AtomicGkw} 
\end{eqnarray}
which is equivalent to the self-energy form
 \begin{eqnarray}
\Sigma(k,\omega)=\frac{U^2}{4}\frac{1}{\omega}.
\label{AtomicSkw} 
\end{eqnarray}
Equation (\ref{AtomicSkw}) is interpreted by the emergence of the pole of $\Sigma(k,\omega)$ at $\omega=0$ independent of momenta $k$.
In fact, this pole generates the Mott gap of the atomic limit and the Mott insulator emerges at half filling.
This is an example where the gap can be generated from a pole of the self-energy even when an apparent spontaneous symmetry breaking is absent.

In the case of the Mott gap, it was pointed out\cite{Zhu-Zhu} that by employing 
\begin{equation}
d^{\rm (MG)}_{i\sigma}=c_{i\sigma}(1-2n_{i-\sigma}),
\label{dMG}
\end{equation}
the interaction term in the Hubbard model can be rewritten as the noninteracting two-component Hamiltonian, containing the hybridization between the two component $c_{i\sigma}$ and $d^{\rm (MG)}_{i\sigma}$.  

Here, we show that a two-component fermion model is \red{exactly} equivalent to the Hubbard model in the atomic limit within the Hilbert space of the atomic Hubbard model.
For that purpose we introduce 
\begin{eqnarray}
\tilde{c}_{i\sigma}&=&c_{i,\sigma} \\
\tilde{d}_{i\sigma}&=&d^{\rm (MG)}_{i\sigma}
\label{tildec} 
\end{eqnarray}
With this two-component fermions one can show that the Hamiltonian
\begin{eqnarray}
H_{cd} &=&\sum_{i\sigma}[ \epsilon_{\tilde{c}}\tilde{c}^{\dagger}_{i\sigma}\tilde{c}_{i\sigma}+\epsilon_{\tilde{d}}\tilde{d}^{\dagger}_{i\sigma}\tilde{d}_{i\sigma} +\Lambda( \tilde{c}^{\dagger}_{i\sigma}\tilde{d}_{i\sigma}+ \tilde{d}^{\dagger}_{i\sigma}\tilde{c}_{i\sigma})] \ \ \ \ \ \
\label{atomicTCFM} 
\end{eqnarray}
with \red{
\begin{eqnarray}
\epsilon_{\tilde{c}}=\epsilon_{\tilde{d}}=-\Lambda=U/2
\label{cde} 
\end{eqnarray}
}
\red{
is equivalent to the Hubbard model in the atomic limit:
\begin{eqnarray}
H_U &=&U\sum_{i}n_{ci\uparrow}n_{ci\downarrow} \ \ \ \ \
\label{eq:HubbardU} 
\end{eqnarray}
}
\red{
After diagonalization of Eq.(\ref{atomicTCFM}), 
the diagonalized state is given by the bonding and antibonding states as 
\begin{eqnarray}
b_{\sigma}&=& \frac{1}{\sqrt{2}}(\tilde{c}_{\sigma}+\tilde{d}_{\sigma})=\sqrt{2}c_{\sigma}(1-n_{-\sigma}), \label{eq:ab0}  \\
a_{\sigma}&=& \frac{1}{\sqrt{2}}(\tilde{c}_{\sigma}-\tilde{d}_{\sigma})=\sqrt{2}c_{\sigma}n_{-\sigma}.
\label{eq:ab} 
\end{eqnarray} 
The bonding and antibonding states are nothing but the lower and upper Hubbard levels, respectively, whose averaged energies are given by $E=0$ and $U$. The lower (upper) Hubbard represents the singly (doubly) occupied particle as one sees the last expressions in Eqs.(\ref{eq:ab0}) and (\ref{eq:ab}). 
}
Then the mapping between Eqs.(\ref{eq:HubbardU}) and (\ref{atomicTCFM}) becomes exact. 
For more details, see Appendix\ref{appendix:AtomicLimit}. 
An important point is that the gap ascribed to the many-body effect can be exactly represented by the noninteracting two-component fermion model as a hybridization gap without any spontaneous symmetry breaking, through the fractionalization of electrons into $\tilde{c}$ and $\tilde{d}$. 

When the ratio $t/U$ becomes nonzero, calculations by cluster extension of the dynamical mean-field theory (cDMFT)~\cite{kotliarCDMFT} support that the pole in the self-energy in the atomic limit survives. However, the pole acquires dispersions~\cite{Sakai2009,Sakai2010}. 

Note that the creation operator of the Mott gap fermion $\tilde{d}^{\dagger}$ is expressed as a linear combination of the doublon creation and ``singlon" creation as $\tilde{d}^{\dagger}_{\sigma} =\frac{1}{2}(b_{\sigma}^{\dagger}-a_{\sigma}^{\dagger}$as an anti-resonant state of the two.  In the strong coupling Hubbard model, the doublon and hole are strongly bound each other as we see below and they form an exciton, where the doublon and hole are resonating and exchanging their positions at the bond of exciton, which are also dynamically fluctuating with a singly-occupied pair of spin singlet (resonating valence bond (RVB)) with the weight proportional to $t/U$ of the RVB weight. Therefore, the Mott gap fermion is born as a fermion component of the tightly bound and bosonic exciton, which will be discussed in detail later in Sec.\ref{Sec2}.

Upon small carrier doping to the Mott insulator, several numerical studies have reproduced that the pseudogap much smaller than the Mott gap opens above the superconducting critical temperature $T_{\rm c}$ near the Fermi level and coexists with the Mott gap~\cite{Kyung,Stanescu,ZhangImada,Sakai2009,Kusunose,Liebsch,Yamaji,YamajiPRB,Sordi,Gull2013,Sadovskii2005}. The coexistence of the pseudogap and the Mott gap is a consequence of the emergence of another self-energy pole in $\Sigma$ in the original Mott gap. 
Although several spontaneous symmetry-broken phases such as the stripe ordering and a flux (or $d$-density wave) state were proposed as the origin of the pseudogap~\cite{Ghiringhelli,Daou,Lawrer,Boeuf,flux}, it is unclear whether such a symmetry breaking is universal and exists in all the pseudogap states in the cuprates.  Since the pseudogap is present as well in the absence of symmetry breaking in the cDMFT studies, we are urged to understand a mechanism working without assuming any symmetry breaking as in the case of the Mott gap.
When the origin of the pseudogap is ascribed to a generic hybridization mechanism, one needs to identify the hidden fermion (dark fermion) object $\tilde{d}$ in Eq.(\ref{atomicTCFM}) that generates the pseudogap.

To gain insight into the nature of the pseudogap formation, it is useful to examine the superconducting phase as well. If the hybridization mechanism works, a natural phenomenological extension of Eq.(\ref{TCfermionNormal}) to the superconducting state is the Hamiltonian\cite{Sakai2016}
\begin{eqnarray}
H&=&\sum_{k,\sigma}[ \epsilon_c (k)c_{k,\sigma}^{\dagger}c_{k,\sigma} +\epsilon_d (k)d_{k,\sigma}^{\dagger}d_{k,\sigma}  \nonumber 
\\
&+& \Lambda (k) (c_{k,\sigma}^{\dagger}d_{k,\sigma} +{\rm H.c.})
\nonumber 
\\
&+&(\Delta_c(k) c_{k,\sigma}^{\dagger}c_{-k,-\sigma}^{\dagger}+\Delta_d(k) d_{k,\sigma}^{\dagger}d_{-k,-\sigma}^{\dagger} + {\rm H.c})
], \nonumber \\
\label{TCfermionAnomalous} 
\end{eqnarray}
where the anomalous part proportional to the superconducting order parameters $\Delta_c(k)$ and $\Delta_d(k)$ becomes nonzero.

By solving Eq.(\ref{TCfermionAnomalous}), Green's function for $c$ particle is obtained as
\begin{eqnarray}
G_c(k,\omega)=\frac{1}{\omega-\epsilon_c(k)-\Sigma^{\rm nor}(k,\omega)-W(k,\omega)},
\label{SCGkw} 
\end{eqnarray}
with 
\begin{eqnarray}
W(k,\omega)=\frac{\Sigma^{\rm ano}(k,\omega)^2}{\omega+\epsilon_c(k)+\Sigma^{\rm nor}(k,-\omega)^*},
\label{SCGkw2} 
\end{eqnarray}
\txr{\begin{eqnarray}
\Sigma^{\rm nor}(k,\omega)=\frac{\Lambda(k)^2(\omega+\epsilon_d(k))}{\omega^2-\epsilon_d(k)^2-\Delta_d(k)^2},
\label{SCGkw3} 
\end{eqnarray}
and 
\begin{eqnarray}
\Sigma^{\rm ano}(k,\omega)=\Delta_c(k)+\frac{\Lambda(k)^2\Delta_d(k)}{\omega^2-\epsilon_d(k)^2-\Delta_d(k)^2}.
\label{SCGkw4} 
\end{eqnarray}
}
Now the pole position of $\Sigma^{\rm nor}$ at \txr{$\omega=\epsilon_d(k)$} in the normal state (expected to generate the pseudogap) is modified to  \txr{$\omega=\pm\sqrt{\epsilon_d(k)^2+\Delta_d(k)^2}$}.
Remarkably, the anomalous part $\Sigma^{\rm ano}$ also has a pole exactly at the same position.  Accordingly, $W$ (Eq.(\ref{SCGkw2})) in the denominator of $G$ in Eq.(\ref{SCGkw}) also has a pole of the order 1 at the same energy. Surprisingly, the residue of the poles of $W$ and $\Sigma^{\rm nor}$are shown to have exactly the same amplitude but the opposite sign\cite{Sakai2016}.
This means that the pole structure in $G$ generating the pseudogap in the normal state has to immediately disappear in the superconducting state.  If the origin of the pseudogap is not ascribed to this hybridization mechanism, such a remarkable cancellation would not be expected. (Actually if the pseudogap arose from the coupling to the boson-like mode such as the spin fluctuation, the cancellation would not happen\cite{Sakai2016}. \txr{See supplementary materials of Ref.\citen{Sakai2016} on the inconsistency with the bosonic glue model, \red{where bosons are coupled essentially in the form of Eq.(\ref{TCfermionboson}). The dynamical coupling to bosons generically introduces convolution in the energy integral and the resultant retardation does not allow instantaneous fermion hybridization, while the instantaneous hybridization is strictly required for the present cancellation to occur. }})  Therefore, the disappearance of the pole structure in $G$ against the pole in each self-energy is a conclusive testimony of the present hybridization mechanism.

In the superconducting phase (at temperature $T<T_{\rm c}$) in the solution of the cDMFT~\cite{Sakai2016}, it was shown that the pole responsible for the pseudogap in the normal state continues and survives in the normal self-energy $\Sigma^{\rm nor}$, while the pole in the anomalous self-energy $\Sigma^{\rm ano}$  
emerges as well. The pole energies of $\Sigma^{\rm nor}$ and $\Sigma^{\rm ano}$ are always the same though the value depends on temperature, doping concentration and interaction strength. In addition, it was found that the poles originated from $\Sigma^{\rm nor}(k,\omega)$ and $W$ perfectly cancel in the sum $\Sigma^{\rm nor}(k,\omega)+W$ in perfect agreement with the expectation from the above two-component hybridization theory.

Furthermore this pole of $\Sigma^{\rm ano}$ at the same position of the pole of $\Sigma^{\rm nor}(k,\omega)$ generates a prominent peak in the imaginary part of $d$-wave superconducting gap function ${\rm Im}\Delta (k,\omega)$ by the relation 
\red{
\begin{eqnarray}
\Delta (k,\omega)=Q(k,\omega)\Sigma^{\rm ano}(k,\omega)
\label{eq:Deltakw}
\end{eqnarray}
 with
\begin{equation}
Q(k,\omega)=
\frac{1}{1-\left[ \Sigma^{\rm nor}(k,\omega)-\Sigma^{\rm nor}(k,-\omega)^{\ast}\right]/(2\omega)}
\mid_{\delta\rightarrow +0}, 
\label{eq:Q}
\end{equation}
where the quasiparticle renormalization factor $z(k)$ is related by the relation
$z(k)=\lim_{\omega\rightarrow 0}Q(k,\omega)$.
}  
Because this prominent peak in ${\rm Im}\Delta (k,\omega)$ contributes to more than 80\% of the real superconducting gap ${\rm Re}\Delta(\omega=0)$ through the Kramers-Kronig relation, it was shown to be the primary origin of the high $T_{\rm c}$\cite{MaierPoilblancScalapino}. 

Then the crucial question for the high-$T_{\rm c}$ mechanism is the physical mechanism of the emergence of the poles of $\Sigma^{\rm nor}$ and $\Sigma^{\rm ano}$ or in other words, the origin of the \red{dark} fermion $d$ because the pole position of $\Sigma^{\rm nor}$ is nothing but the bare dispersion of $d$ in the hybridizing Hamiltonian.
 
In this paper, we examine a possible physical object of the \red{dark} fermion. 
One of the authors proposed before that the origin of the \red{dark} fermion is the quasiparticle bound to a hole in the underdoped Mott insulator~\cite{Yamaji,YamajiPRB} by using the slave boson formalism.
In this paper, we formulate such a \red{dark} fermion tightly bound to the exciton, by gaining insight from the mass generation mechanism of hadrons and quarks in high energy physics and by comparing with other gap generation mechanism in condensed matter physics including the Mott gap generation.

In Sec.\ref{Sec2}, we discuss the role of exciton in the Mott insulator. In Sec.\ref{Sec3}, we propose the nature of the \red{dark} fermion as a composite fermion.  In Sec.\ref{Sec4}, we discuss the origin of the pseudogap in terms of of the \red{dark} fermion. In Sec.\ref{Sec5}, we present a formalism to study relevant fermionic excitations including the above \red{dark} fermion.
Section \ref{Sec6}  is devoted to discussions about the relation to other numerical and experimental studies. 

\section{Exciton in the Mott insulator}\label{Sec2}
In the Mott insulator of the Hubbard model at half filling, the doubly occupied site (doublon) represented by $n_{i,\uparrow}n_{i,\downarrow}=1$ and the empty site (holon) $(1-n_{i,\uparrow})(1-n_{i,\downarrow})=1$ form a bound state with the binding energy of the order of $U$, if they are nearby. This excitation is identified as a locally bound Frenkel-type exciton. The exciton dynamics was discussed in a context \red{quite} different from the present study~\cite{Wrobel-Eder,Rademaker_Zaanen_exciton} \red{Here, we discuss a novel and crucial role of excitons in physics of doped Mott insulators by focusing on the connection to the mechanism of superconductivity and pseudogap formation  from a general perspective}.
If $t/U$ is nonzero, the density of such excitons is nonzero even in the ground state.  The creation operator of the exciton in the Mott insulator can be written as 
\begin{eqnarray}
b_{j,\delta}^{\dagger}&\equiv & B\sum_{\sigma}
c_{j,\sigma}^{\dagger}c_{j+\delta,\sigma} n_{j,-\sigma}(1-n_{j+\delta,-\sigma}),
\label{b} 
\end{eqnarray}
with a normalization constant $B$ to ensure the bosonic commutation relation for $b$ and $b^{\dagger}$. \red{Note that this is the lowest order process of the exciton generation in terms of $t/U$ expansion (strong coupling expansion), because the upper and lower Hubbard band particles are described by Eqs.(\ref{eq:ab0}) and (\ref{eq:ab}).}
The operator $b^{\dagger}$ represents a part of the kinetic energy $c_{j,\sigma}^{\dagger}c_{j+\delta,\sigma}$ and $b^{\dagger}_{j,\delta}$ behaves as creating a boson when the doublon at the site $j$ and the holon at the site $j+\delta$ are bound. Although the creation energy (with a dispersion) of $b$ is high in the order of $U$ above the Fermi level, the quantum fluctuation generated by the transfer term in the Hubbard model generates a finite density of excitons even in the ground states in contrast to the conventional band insulators and semiconductors in the noninteracting limit. In other words, the vacuum of the Mott insulator can be regarded to have a real nonzero fluctuation (polarization) generating excitons. 
  
In the Mott insulator, the charge (single-particle) degrees of freedom are gapped, while the excitons are fluctuating and dynamical in addition to the spins.   Although the spins have been well studied with their antiferromagnetic long-range order or its strong fluctuations, the role of excitons is not well understood.  Since the exciton density is finite and dynamical even in the ground state of the Mott insulator, an effective Hamiltonian for the exciton can be derived in a similar way to the derivation of the Heisenberg model in the case of the spins.  However, the dynamics of the creation or annihilation of excitons is described just by the first order electron hopping process in contrast to the second-order perturbation in terms of $t/U$ needed in the Heisenberg exchange interaction for the spins to emerge.  The motion of created exciton is generated by the second order process.

More concretely, in the effective Hamiltonian for the exciton, the electron hopping $t_{\delta}$ in the Hubbard model generates the exciton creation/annihilation from the ground state as
\begin{eqnarray}
H_{b}^{(0)} &= & -\sum_{{j}, \delta } \lambda_{\delta}\left[b^{\dagger}_{{j},\delta}+ b_{{j},\delta}
\right],
\label{Hb0}
\end{eqnarray}
where the summation over $\delta$ represents the form and the extension of the exciton in Eq.(\ref{b}), and ${j}$ is the exciton site represented by the doublon site. The amplitude $\lambda_{\delta}$ should be \red{trivially} proportional to $t_{\delta}$. 
\red{Since the exciton creation energy is proportional to $U$, the second order process proportional to $t_{\delta}^2/U$ yields the recombination of the exciton into singly occupied states.}

The number of excitonic bound states may depend on $U/\lambda_{\delta}$ and here the number \red{will be} denoted by $N_b$. \red{The binding energy of the exciton is defined by the energy difference from the formation energy of isolated one doublon and one holon far apart at infinite distance.} The noninteracting part of the diagonalized exciton effective Hamiltonian reads
\begin{eqnarray}
H^{\rm nonint}_{\tilde{b}} & = & H_{\tilde{b}}^{(0)} +H_{\tilde{b}}^{(1)}
\label{HbnointD}
\end{eqnarray}
with
\begin{eqnarray}
H_{\tilde{b}}^{(0)}  & = &  \sum_{j}\sum_{l=1,N_b} \zeta_l 
\left[ \tilde{ b}_{{j}l}^{\dagger}+\tilde{b}_{{j}l}\right],
\nonumber \\
H_{\tilde{b}}^{(1)}  & = & \sum_k\sum_{l=1,N_b} \epsilon_b^{(l)}(k)\tilde{b}_l^{\dagger}({k})\tilde{b}_l({k}),
\label{HbnointD01}
\end{eqnarray}
where 
\begin{eqnarray}
\tilde{b}_{l}(k)=\sum_{j=1,N} \tilde{b}_{\txr{j l}}e^{ikj}/\sqrt{N}.\label{bk}
\end{eqnarray}
 Here, $N$ is the number of sites and 
\begin{eqnarray}
\tilde{b}_{{j}l}&=&\sum_{\delta} {\mathcal U}_{l\delta}b_{{j},\delta},  \nonumber \\
\zeta_l&=&\sum_{\delta} {\mathcal U}_{l\delta}^{-1}\lambda_{{j},\delta}
\label{b-unitary}
\end{eqnarray}
with
the unitary transformation $\mathcal U$ diagonalizing the exciton dispersion Eq.~(\ref{HbnointD01}) with respect to $l$. \red{We do not need to determine the form of the dispersion for the later discussion, while it is a well defined quantity and can be straightforwardly calculated from $\langle \Phi_0 |\tilde{b}_l({k}) P_{\Phi_0} H_{\rm Hub}  P_{\Phi_0}\tilde{b}_l^{\dagger}({k}) |\Phi_0 \rangle/\langle \Phi_0 |\tilde{b}_l({k})P_{\Phi_0} \tilde{b}_l^{\dagger}({k}) |\Phi_0 \rangle$ after determining the variational form of  $\tilde{b}_l^{\dagger}({k}) |\Phi_0 \rangle$, where $P_{\Phi_0}=1-|\Phi_0\rangle\langle\Phi_0|/\langle\Phi_0|\Phi_0\rangle$ is the projection operator to construct states orthogonal to the ground state. It should be noted anyhow $\epsilon_b^{(l)}(k)$ has the energy scale of $U$ as we already mentioned.}
 
The ground state of this Hamiltonian is exactly given by  
\begin{eqnarray}
|\Phi_0^{\rm nonint}\rangle =\prod_{j}^{N} \prod_l^{N_b}\left[ 1- \frac{\zeta (\tilde{b}_{jl})^\dagger}{\epsilon_b^{(l)}(k=0)}\right]|0\rangle,
\label{HbGS}
\end{eqnarray}
where $|0\rangle$ is the vacuum of the exciton state derived from a ground state of the Hubbard model at half filling in the atomic limit.

\begin{figure}[h!]
  \begin{center}
    \includegraphics[width=7cm]{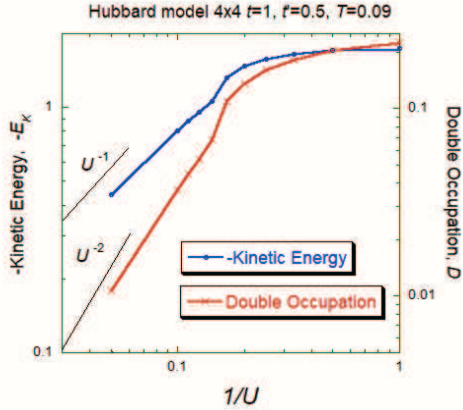}
    \end{center}
\caption{
$t/U$ dependence of  doublon density and kinetic energy per site for an example of the square-lattice Hubbard model at half filling. In this example, the next neighbor hopping is $t'=0.5$ in the energy unit of the nearest neighbor transfer $t=1$. The calculation was performed by using a method~\cite{Imada1986,Sugiura} for finite temperatures. 
 The data were obtained by using the H$\Phi$ code~\cite{HPhi,HPhi2}  for 4 by 4 lattice with the periodic boundary condition at temperature $T=0.09$,  which already shows convergence to the ground state~\cite{MisawaYamaji}. 
}
\label{doublon_density}
\end{figure}

\begin{figure}[h!]
  \begin{center}
    \includegraphics[width=7cm]{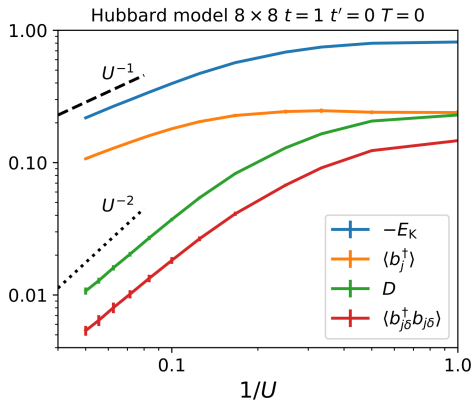}
    \end{center}
\caption{
\red{$t/U$ dependence of  kinetic energy $-E_K$, doulon density $D$, condensation amplitude $\langle b_j\rangle=\langle b^{\dagger}_j\rangle$ and $\langle b^{\dagger}_{j\delta}b_{j\delta}\rangle$ (with the nearest neighbor pair at $j+\delta$ and $j$) per site 
for an example of the ground state of $8\times 8$ square-lattice Hubbard model at half filling with the antiperiodic-periodic boundary condition. 
In this example, only the nearest neighbor hopping $t$ is nonzero and taken as the energy unit as $t=1$. The calculation was performed by the variational Monte Carlo method~\cite{Tahara-Imada,mVMC,mVMC2}. We employ the Gutzwiller-Jastrow factors, doublon-holon correlation factors, and the generalized pairing wave function with $2\times 2$ sublattice structures. The error bars indicate the estimated statistical errors of the Monte Carlo sampling.}
}
\label{exciton_density}
\end{figure}

\begin{figure}[h!]
  \begin{center}
    \includegraphics[width=7cm]{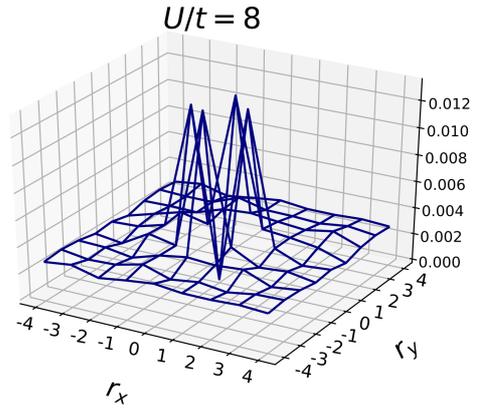}
    \end{center}
\caption{
\red{Spatial correlation of the doublon and holon $C_{dh}(\vec{r}=\vec{r}_i-\vec{r}_j)=\langle n_{i\sigma}n_{i-\sigma}(1-n_{j\sigma})(1-n_{i-\sigma})\rangle$ for the Mott insulator at $U/t=8$ for $8 \times 8$ Hubbard model with only the nearest neighbor hopping $t$ and taken as the energy unit as $t=1$. It indicates that they are strongly bound mostly at the nearest neighbor site. }
}
\label{doublon-holon-correlation}
\end{figure}

In contrast to the ordinary coupled electron-phonon or electron-photon systems, here the Bose-Einstein condensation of excitons occurs with macroscopic concentration of bosons. 
The condensation amplitude is given by
$\langle \tilde{b} \rangle =\zeta/\epsilon^{(l)}_b({k}=0)$.
Since $\zeta$ is scaled by $t_{\delta}$, the condensation has the amplitude $\langle \tilde{b} \rangle$ scaled by $t_{\delta}/U$, which is the same as the averaged kinetic energy obtained from the first term in the right hand side of Eq.~(\ref{Hubbard}). \txr{((Eq.(\ref{b}) contains a constraint so that the initial states of the two sites involved in the electron transfer are both singly occupied in contrast to the kinetic energy term in the Hubbard Hamiltonian. However, such a constraint is satisfied in most sites of the Mott insulator and lightly doped Mott insulator. $U$ dependences of the kinetic energy ($\propto 1/U$) is shown in Fig.\ref{doublon_density}. \red{The same scaling of $\langle b_{i\delta}\rangle \propto 1/U$ is shown in comparison of Figs.\ref{doublon_density} and \ref{exciton_density})}}. 
The exciton density is scaled by $\langle \tilde{b}_{i\delta}^{\dagger}\tilde{b}_{i\delta} \rangle \propto (t_{\delta}/U)^2$ and is the same as the doublon or holon density $\langle n_{i\uparrow}n_{i\downarrow}\rangle$. \red{The same scaling of $\langle b_{i\delta}^{\dagger}b_{i\delta}\rangle \propto 1/U^2$ and the doulon density is confirmed in Figs.\ref{doublon_density} and \ref{exciton_density}}. 
\txr{(Note that the doublons are bound to the holons in pair as the excitons in the Mott insulator and therefore the density of doublon or holon is \red{essentially} the same as the exciton density \red{as one can see in Fig.\ref{exciton_density}.) See also the spatial correlation of the doublon and hole in Fig.\ref{doublon-holon-correlation}, which shows that the doublon and hole are bound mostly to the nearest neighbor site in the Mott insulator at large $U/t$.}} 

The difference from the conventional electron-phonon and electron-photon coupled systems is that the ``symmetry breaking field" represented by $H_b^{(0)}$ is present and the boson density is nonzero even in the ground state, if $t_{\delta}/U$ is nonzero. Namely, this condensation is not the consequence of spontaneous symmetry breaking.   
Although the single exciton energy level is high in the order of $U$, $H_b^{(0)}$ violating the exciton-number conservation introduces the quantum fluctuation and generates the condensation.

Using the variational Monte Carlo method, we have also calculated the ground-state average of the Fourier transform Eq.~({\ref{bk}) with the summation over $\delta$ in Eq.(\ref{b-unitary}) only for the nearest neighbor site under the assumption about the number of the bound state $N_b=1$.  We assume that a constant $\mathcal{U}_{l=1,\delta}=u$ satisfies the isotropic $s$-wave symmetry.  The result is plotted in Fig.~\ref{BEC} for the Hubbard model on the square lattice with only the nearest neighbor transfer $t_{\delta}=1$ and $U=10$ for $8\times 8$ lattice. The delta function peak of $\langle \tilde{b}_{k=0}\rangle=\langle \tilde{b}_{k=0}^{\dagger}\rangle$ indicates the Bose-Einstein condensation of the exciton. Here the exciton is assumed as the nearest neighbor pair of  the doublon and holon.
\begin{figure}[h!]
  \begin{center}
    \includegraphics[width=6cm]{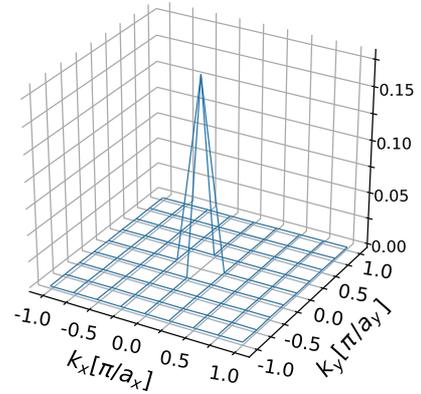}
    \end{center}
\caption{
The exciton condensation amplitude $\langle \tilde{b}_{k}\rangle=\langle \tilde{b}_{k}^{\dagger}\rangle$ in the Brillouin zone of the momentum $k$ for the square lattice Hubbard model with the nearest neighbor transfer $t_{\delta}=1$ at $U/t_{\delta}=10$ at half filling.  Here $a_x=a_y=1$ is the lattice constant. The calculation was performed by using a multi-variable variational Monte Carlo method~\cite{Tahara-Imada,mVMC,mVMC2} discussed in Sec.\ref{Sec5}.
}
\label{BEC}
\end{figure}
    
The exciton has a local nature and has a hard core. In addition, the excitons are polarized and interact through the dipole interaction. We describe these interactions as 
\begin{eqnarray}
H_{b}^{\rm int}  & = & \sum_{{i},{j}}\sum_{l,l'} V_{{i},{j}}n_{{b}_{{i}l}}n_{{b}_{{j}l'}}.
\label{Hbint}
\end{eqnarray}
We do not go into details of the interaction at this stage.
However, such interactions introduce the higher order terms in the Landau-Ginzburg expansion of the exciton, which leads to the Gross-Pitaevskii-type Hamiltonian given by 
\begin{multline}
H_{b}^{\rm eff}=  
\zeta_1 (\langle \tilde{b}_1\rangle+\langle \tilde{b}_1\rangle^* ) +\epsilon_b^{(1)}(k=0)|\langle \tilde{b}_1\rangle|^2
\\
+\sum_k \frac{k^2}{2}\frac{d^2\epsilon_b^{(1)}(k)}{dk^2}|_{k=0} |\langle \tilde{b}_1\rangle|^2 + \sum_{{ i},{j}}V_{{i},{j}}|\langle \tilde{b}_1\rangle|^4 
\label{Hbeff}
\end{multline}
for small $k$.  Here we assumed only one bound state for the exciton ($N_b=1$) in the insulator for simplicity and $\langle \tilde{b}_1\rangle\equiv \langle \tilde{b}_{j1}\rangle$ is its uniform condensate amplitude.
It enhances the amplitude of $\langle \tilde{b}_1\rangle$ when the quadratic term coefficient becomes negative, which is the ordinary route of the U(1) gauge symmetry breaking in the Bose-Einstein condensation. 

We here note the relation of the exciton to the fermionic excitation. An electron (hole) constituting an exciton yields (after the breakup of the exciton), the upper (lower) Hubbard band.
However, such a breakup of the exciton is absent in the ground state of the Mott insulator because of the confinement and is seen only as high-energy excitations as the upper and lower Hubbard bands. Note also that the creation of a particle $d^{\rm (MG)\dagger}$ represents essentially the anti-phase linear combination of an electron (lower Hubbard electron) added to a hole state (included in the exciton) and to a singly occupied site (upper Hubbard electron) as
$d_{\sigma}^{\rm \dagger (MG)}=c_{\sigma}^{\dagger}(1-n_{-\sigma})-c_{\sigma}^{\dagger}n_{-\sigma}$ (see Eq.(\ref{dMG})).

\section{Electron (hole) bound to holon (doublon) as composite particles in the doped Mott insulator -- dark fermion}\label{Sec3}
When carriers are doped into the Mott insulator, the mutual screening weakens the binding potential of the doublon and holon
and the mean distance of the bound doublon and holon (in the exciton) is expected to increase because of the increasing itinerancy.  In this circumstance, because of the increase of the spatial extension of binding interaction range, the number of exciton bound level may increase with the addition of the Wannier-type excitonic states, (which is represented by $\tilde{b}_{jl}$ with $l\ge 2$ in the notation of Eq.(\ref{HbnointD01})),
which may have excitation energies much smaller than the Mott gap when the doping concentration increases.

Although the exciton is a bosonic excitation, a fermion called dark fermion can be generated from this Wannier-type exciton by its breakup. In analogy with the Mott gap fermion, this dark exciton is represented by the linear combination of spatially extended doublon creation operator at the singly occupied site and the ``singlon" creation operator at the hole site. This dark fermion constitutes an ingap state 
represented by the hole (particle)-type composite fermion if it becomes unbound into an electron, which may be detected as the ingap peak in the optical conductivity~\cite{takagi}. 

The creation of a composite particle  (\red{dark} fermion)
can be variationaly written as
\begin{eqnarray}
d_{j\sigma}^{\dagger}&=&\sum_{\delta} d_{j,\delta,\sigma}^{\dagger}
 \ \ \ \ \ \ 
\label{eq:d_cp}
\end{eqnarray} 
and
\begin{multline}
d_{j,\delta,\sigma}^{\dagger}\equiv G_{j,\delta,\sigma}^{\rm (DP)}c_{j,\sigma}^{\dagger} \\
G_{j,\delta,\sigma}^{\rm (DP)} \equiv g_{\delta}^{\rm (DP)}
-\alpha_{\delta}^{\rm (DP)}n_{j+\delta,-\sigma}-\beta_{\delta}^{\rm (DP)}n_{j+\delta,\sigma} \\
+ \gamma_{\delta}^{\rm (DP)}n_{j+\delta,\sigma}n_{j+\delta,-\sigma}.
\label{d}
\end{multline}
where we expect $g_{\delta=0}^{\rm (DP)} \sim 1$, $\alpha_{\delta=0}^{\rm (DP)}$ comparable to 2, and other parameters small, 
similarly to $d^{\rm MG \dagger}$ in Eq.(\ref{dMG}), but the nonzero parameters at nonzero $\delta$ is important to reflect the extension of the associated Wannier exciton distinct from the Mott gap exciton (Frenkel exciton).
Numerically, $g_{\delta}, \alpha_{\delta}$, $\beta_{\delta}$ and $\gamma_{\delta}$ (see Fig.~\ref{fig:Fig1}) are variational parameters to be determined later.

Here, the variational ground state wavefunction $|\Phi_0\rangle$ has to be given beforehand. 
We note that the form (\ref{eq:d_cp}) can represent the particles for the Mott gap $d_{j\sigma}^{\rm (MG) \dagger}$ and the quasiparticle as well, as we discuss later
.
\begin{figure}[h!]
  \begin{center}
    \includegraphics[width=3cm]{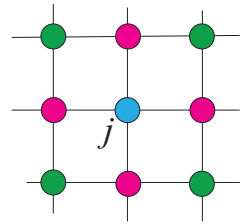}
    \end{center}
\caption{
Schematic illustration of range of $g_{\delta}, \alpha_{\delta}, \beta_{\delta}$ and $\gamma_{\delta}$ in Eq. (\ref{d}) and $\mathcal{U}_{l\delta}$ in Eq.(\ref{b-unitary})  in case of square-lattice Hubbard model.  $j$ is at the center and $g_{\delta}$ etc. may depend on the colors of site $j+\delta$ sites. 
(Sites beyond the next nearest neighbor $\delta$ are omitted,
but should be considered in better estimates).
}
\label{fig:Fig1}
\end{figure}

The dispersion of the \red{dark} particle $d$ is determined from 
the Fourier transform of 
\begin{eqnarray}
t^{(d)}_{i-j}&=&\langle \Psi_{\rm DP}(i) |H|\Psi_{\rm DP}(j)\rangle /\langle \Psi_{\rm DP}(i) |\Psi_{\rm DP}(j)\rangle 
\label{eq:t(d)}
\end{eqnarray}
where $\langle \cdots \rangle$ expresses the average over the ground state $|\Phi_0\rangle$ at half filling and can be estimated by an accurate estimate of the ground state wavefunctions, for example, by the variational Monte Carlo or tensor network method\cite{Tahara-Imada,Zhao2017}.
The number of bound states of Wannier-type exciton generated upon doping can be larger than one, but in the following discussion, we assume just one Wannier-type state for simplicity in addition to the Frenkel-type exciton.

Even in the doped case, the strongly bound Frenkel exciton (and resultant upper and lower Hubbard bands) may survive in the underdoped region, although it disappears in the overdoped region as well as the Wannier-type bound states.
In addition, the quasiparticle extended in space independently appears as an electron (a hole) unbound from holes (doublons). Therefore we consider three types of fermionic excitations from the ground state of the doped Hubbard model in the particle- and hole- excitation side each. One is the quasiparticle (quasihole) excitation and the second is a \red{dark} fermion composed of an electron (hole) trapped around a hole (doublon). These two types of excitations exist only in the doped Mott insulator and do not exist in the Mott insulator.
The third is an electron (hole) extracted from doublon (hole) in the Frenkel exciton and it constitutes upper (lower) Hubbard band. 

We expect that the quasiparticle gains the kinetic energy because of its spatially extended nature, while the \red{dark} fermion may also have a low energy and at a local minimum as a bound state because the onsite Coulomb energy is lowered if the electron stays at a hole. These two states then are in competition and both may appear near the Fermi level as low-energy excitations. Their energies may depend on the doping concentration and momentum; the \red{dark} fermion may have an energy lower than the quasiparticle when the doping concentration becomes small, because the scale of the kinetic energy becomes smaller than the interaction energy, when the system approaches the Mott insulator.   
 
\section{Pseudogap Generation}\label{Sec4} 
A relevant question is whether the effective hamiltonian contains the instantaneous hybridization term $c^{\dagger}_{i\sigma}d_{j\sigma}+d^{\dagger}_{j\sigma}c_{i\sigma}$ as the process derivable from the original Hubbard model (or any other theoretical model for the cuprate superconductors), if $d$ is the proposed \red{dark} fermion, similarly to the case of Mott gap \red{fermion}. Namely, in the present candidate, $d$ represents an electron weakly bound to a holon, which constitutes a Wannier type exciton. There we expect that the holon and the electron are apart with a substantial distance \red{in comparison to the Frenkel exciton}. 

In the Mott insulator, the above term can be generated directly in the first order process only when the original hamiltonian contains the long-ranged electron hopping term. If the hopping is restricted to short-ranged pairs such as the nearest neighbor sites, the first order process primarily generates only the Mott exciton (Frenkel exciton) as Eq.(\ref{b}) and associated Mott gap particle $d^{\rm (MG)}$.  In other words, the density of \red{Wannier} excitons and the associated \red{dark} fermions is zero.  

If the holes are doped, however, a quasiparticle may be trapped to the existing doped hole, while such a process happens only when doped holes, quasiparticles, and the Wannier-type excitonic bound state exist.   Since the Frenkel excitons exist already in the Mott insulator, even in the limit of dilute hole density, the \red{dark} fermion $d$ can still be generated from the Frenkel exciton $\tilde{b}_1$ in a single-particle hopping process
$t_{n-j}d^{\dagger}_{n,\sigma}c_{j+\delta,\sigma}\tilde{b}_{j1}$,
where a Frenkel-type exciton, bearing the component consisting of the doublon at the site $j$ and the holon at the site $j+\delta$, is annihilated and an unbound holon is recreated at the site $j+\delta$, together with the creation of a \red{dark} fermion at the site $n$.  

This process is caused by the hopping of an electron from the site $j$ to $n$.
Thanks to the condensation part of $\tilde{b}_1$,
it generates an instantaneous term
$t_{n-j}d^{\dagger}_{n,\sigma}c_{j+\delta,\sigma}\langle \tilde{b}^{\dagger}_{j1}\rangle$,
which is nothing but the hybridization term between $c$ and $d$.
This is in sharp contrast with the conventional electron-phonon or electron-photon systems,
where Eq.(\ref{TCfermionboson}) does not generate Bose-Einstein condensation,
since the phonons and photons disappear in the ground states in the normal condition in contrast to the real vacuum fluctuation in the Mott insulator.
Even in the absence of the BEC, the bosons mediate the pairing as in the conventional weak-coupling BCS mechanism and the mass generation of quarks, while the strong-coupling superconductivity requires the instantaneous 
hybridization of fermions as we observed in the cDMFT study, where particle-hole bound states (excitons) generate the electron fractionalization into quasiparticles and \red{dark} fermions and their mutual hybridization causes the hybridization gap, one is the Mott gap and the other is the pseudogap.
Superconductivity, a consequence of the particle-particle bound state (Cooper pair), is then boosted up by the \red{dark} fermions.

In the case of the pseudogap, the hybridization term is scaled by $t^2/U$, because $\langle b \rangle=\langle b^{\dagger} \rangle$ is scaled by $t/U$. Then the resultant hybridization gap (namely the pseudogap) is scaled by $t^2/U$, which is much smaller than the Mott gap scaled by $U$.

We note that the hybridization gap proportional to $t\langle b\rangle \propto t^2/U$ stays a constant even in the limit of small doping concentration in agreement with the experimental indication for the pseudogap energy, which increases but saturates with decreasing doping concentration.  This is made possible because of the condensation of $b$ already present in the Mott insulator. 
 
\section{Basic Framework for Numerical Study}\label{Sec5}
For the ground state, we can employ a variational wavefunction obtained, for instance, from variational Monte Carlo calculations, denoted as $|\Phi_0\rangle$. Details of the variational Monte Carlo method are found in Ref.~[\citen{Tahara-Imada}] and other improved possibilities are found in Refs.~[\citen{Zhao2017,Nomura2017}].

\subsection{Particle Excitations}
\subsubsection{Quasiparticle Excitation}

The bare electron with spin $\sigma$ and momentum $k$ added to the ground state is 
\begin{eqnarray}
|\Psi_{\rm e}(k)\rangle &=& c_{k\sigma}^{\dagger} |\Phi_0\rangle, 
\label{eq:electron}
\end{eqnarray}
where the creation operator for an electron with spin $\sigma$ and momentum $k$ is denoted by $c_{k\sigma}^{\dagger}$. The bare excitation energy of an electron is given by
\begin{eqnarray}
E_{\rm e}(k)&=&\frac{\langle \Psi_{\rm e}(k)|H|\Psi_{\rm e}(k)\rangle}{\langle \Psi_{\rm e}(k)|\Psi_{\rm e}(k)\rangle}, 
\label{eq:Ee}
\end{eqnarray}
where $H$ is the Hubbard hamiltonian.

The low-energy single-particle excitation is represented by a quasiparticle after the renormalization arising from many-body effects.
We may assume the real-space operator as
\begin{eqnarray}
|\Psi_{\rm QP}(j)\rangle &=& \tilde{c}_{j\sigma}^{\dagger} |\Phi_0\rangle, 
\label{eq:Psi_qp} \\
\tilde{c}_{j\sigma}^{\dagger}&=&c_{j\sigma}^{\dagger}h+\sum_{\delta} u_{\delta}n_{j+\delta-\sigma} c_{j\sigma}^{\dagger}
\label{tilde_cj}
\end{eqnarray}
to take into account primarily the local correlation, 
where $h_j$ and $u_{\delta}$ (see Fig.~\ref{fig:Fig1}) are variational parameters to be determined. Here we assumed the translational invariance of $u_{\delta}$.
Here and hereafter all the excited state such as $|\Psi_{*}(k) \rangle$ are taken to satisfy the normalization condition $\langle \Psi_{*}(k)|\Psi_{*}(k) \rangle = 1$. 

However, here we assume more general form similarly to
Eqs.(\ref{eq:d_cp}) and (\ref{d}) as
\begin{eqnarray}
\tilde{c}_{j\sigma}^{\dagger}&=&\sum_{\delta} \tilde{c}_{j,\delta,\sigma}^{\dagger}.
 \ \ \ \ \ \ 
\label{eq:c_qp}
\end{eqnarray} 
and
\begin{multline}
\tilde{c}_{j,\delta,\sigma}^{\dagger}\equiv G_{j,\delta,\sigma}^{\rm (QP)}c_{j,\sigma}^{\dagger} \\
G_{j,\delta,\sigma}^{\rm (QP)} \equiv g_{\delta}^{\rm (QP)}
-\alpha_{\delta}^{\rm (QP)}n_{j+\delta,-\sigma}-\beta_{\delta}^{\rm (QP)}n_{j+\delta,\sigma} \\
+ \gamma_{\delta}^{\rm (QP)}n_{j+\delta,\sigma}n_{j+\delta,-\sigma}.
\label{cqp}
\end{multline}
Although the variational form is taken to be the same between the \red{dark} particle and quasiparticle, after the variational determination without bias, $g_{\delta}^{\rm (QP)}$ is expected to be dominant and $\alpha_{\delta}^{\rm (QP)}, \beta_{\delta}^{\rm (QP)}$ and $\gamma_{\delta}^{\rm (QP)}$ are small parameters in contrast to the \red{dark} particle. 

After the Fourier transform, the quasiparticle state may be given by
\begin{eqnarray}
|\Psi_{\rm QP}(k)\rangle&=&\tilde{c}_{k\sigma}^{\dagger} |\Phi_0\rangle=\frac{1}{\sqrt{N}}\sum_{j} \exp[ikr_j]|\Psi_{\rm QP}(j)\rangle, \nonumber \\
\label{eq:Psi_qp_k}
\end{eqnarray}
where the creation operator of the quasiparticle excitation with spin $\sigma$ and momentum $k$ is denoted by $\tilde{c}_{k\sigma}^{\dagger}$.

\subsubsection{\red{Dark} Particle}
We assume that a \red{dark} fermion representing a bound state of an electron with a hole localized at the $j$-th site~\cite{Yamaji,YamajiPRB,Sakai2009,Sakai2010,Imada} is described by Eqs. (\ref{eq:d_cp}) and (\ref{d}).

We then obtain dark particle (composite particle) wave function
\begin{eqnarray}
|\Psi_{\rm DP}(j)\rangle &=& d_{j\sigma}^{\dagger}|\Phi_0\rangle, 
\label{eq:Phi_DP}
\end{eqnarray}
and 
\begin{eqnarray}
|\Psi_{\rm DP}(k)\rangle&=&\frac{1}{\sqrt{N}}\sum_{j} \exp[ikr_j]|\Psi_{\rm DP}(j)\rangle. 
\label{eq:Phi_DP_k}
\end{eqnarray}
\subsubsection{Mott Gap Particle}
The Mott gap particle $d^{\rm (MG)}_{i\sigma}$ adiabatically connected to Eq.~(\ref{dMG}) in the atomic limit can also be represented in the form of Eqs. (\ref{eq:d_cp}) and (\ref{d}), with more localized nature than the \red{dark} particle and $\alpha_{\delta}$ close to 2 in the strong coupling region. Then we seek for
\begin{eqnarray}
|\Psi_{\rm MG}(j)\rangle &=& d_{j\sigma}^{\rm (MG)\dagger}|\Phi_0\rangle, 
\label{eq:Phi_uH}
\end{eqnarray}
with 
\begin{eqnarray}
d_{j\sigma}^{\rm (MG) \dagger}&=&\sum_{\delta} d_{j,\delta,\sigma}^{\rm (MG) \dagger}.
 \ \ \ \ \ \ 
\label{eq:d_MG}
\end{eqnarray} 
and
\begin{eqnarray}
d_{j,\delta,\sigma}^{\rm (MG) \dagger}&\equiv & G_{j,\delta,\sigma}^{\rm (MG)}c_{j,\sigma}^{\dagger} \nonumber \\
G_{j,\delta,\sigma}^{\rm (MG)} &\equiv & g_{\delta}^{\rm (MG)}
-\alpha_{\delta}^{\rm (MG)}n_{j+\delta,-\sigma}-\beta_{\delta}^{\rm (MG)}n_{j+\delta,\sigma} \nonumber \\
&+& \gamma_{\delta}^{\rm (MG)}n_{j+\delta,\sigma}n_{j+\delta,-\sigma}.
\label{dMG2}
\end{eqnarray}
Its Fourier transform $|\Psi_{\rm (MG)}(k)\rangle$ is given by
\begin{eqnarray}
|\Psi_{\rm MG}(k)\rangle&=&\frac{1}{\sqrt{N}}\sum_{j} \exp[ikr_j]|\Psi_{\rm MG}(j)\rangle. 
\label{eq:Psi_dMG_k}
\end{eqnarray}

\subsubsection{Variational Determination of Orthonormalized Excitations}
Because all the excitations are given in the same variational form, we obtain 
the quasiparticle, the \red{dark} particle and the Mott gap excitation at the same time 
in the following way:
First, Calculate $3\times 3$ matrix $\mathcal{N}_{ij}(k,k')= \langle \Phi_{0}|(a^{(i)}(k)a^{(j)\dagger}(k')+a^{(j)\dagger}(k')a^{(i)}(k))|\Phi_{0}\rangle$ with the definition
\begin{eqnarray}
 a^{(1)^{\dagger}}(k)&=& \tilde{c}^{\dagger}(k) \nonumber \\
 a^{(2)^{\dagger}}(k)&=& d^{\dagger} \label{a123} \\
 a^{(3)^{\dagger}}(k)&=& d^{\rm (MG)\dagger} 
\nonumber  
\end{eqnarray}
by assuming some initial conditions of the variational parameters.
(Note that we now employ a set of  variational parameters $g_{\delta}^{\rm (A)}, \alpha_{\delta}^{\rm (A)}, \beta_{\delta}^{\rm (A)}$ and $\gamma_{\delta}^{\rm (A)}$ for $\rm A=QP, DP, MG$.)
In the case of the Mott insulator, only two excitations instead of three in Eq.(\ref{a123}) are sufficient \red{for the description of} the basic dispersion of the upper Hubbard band.

Orthonormalization (diagonalization and normalization) of the matrix $\mathcal N(k)$ by the unitary transformation $P(k)$ and normalization give $\tilde{\mathcal N}(k,k')=P(k){\mathcal N}(k,k')P(k)^{-1}=\delta_{ij}\delta(k,k')N_i(k)$ with the normalized eigenvector $|\tilde{\Psi}_i(k)\rangle\equiv \sum_jP_{ij}(a^{(j)\dagger}(k)+a^{(j)}(k))|\Phi_{0}\rangle/\sqrt{N_i(k)}$ (Namely, $\langle \tilde{\Psi}_i(k)|\tilde{\Psi}_j(k')\rangle=\delta_{i,j}\delta(k,k')$,
if the ground state $|\Phi_0\rangle$ is given in a sector of fixed particle number (canonical ensemble).)

Then by a unitary transformation $U$ of the 3 component basis to diagonalize the energy matrix as $f^{\dagger}=Ua^{\dagger}$, the normal mode for the particle excitation $f^{\dagger}$ satisfies
\begin{eqnarray}
E_{ij}(k)&=&\langle \Psi_{i}(k)|H|\Psi_{j}(k')\rangle=E_i\delta_{ij}\delta(k,k'), \label{eq:Eij}
\end{eqnarray}
with the particle-type elementary excitation
$|\Psi_i(k)\rangle = f^{\dagger}_i(k)|\Phi_0\rangle$, 
where $f$ satisfies the anticommutation relation
$\langle \Phi_{0}|(f^{(i)}(k)f^{(j)\dagger}(k')+f^{(j)\dagger}(k')f^{(i)}(k))|\Phi_{0}\rangle=\delta_{ij}\delta(k,k')$
and $E_i$ is nonnegative by definition because of the nature of particle excitation.

The variational parameters may be determined to lower the lowest energy eigenvalue $E_1$. However, there may be other ways of optimization of the excitations. 
For instance, to extract the physical picture of the Mott gap fermion and composite fermion (\red{dark} fermion), one can alternatively first take $\zeta_{\delta}^{\rm (QP)}=0$ or $\zeta_{\delta}^{\rm (QH)}=0$, for $\zeta=\alpha, \beta, \gamma$, which is the bare electron/hole.
The other excitations can be obtained so that they are orthogonal to it at each momentum.  

Note that the determination of the variational parameters can be done only by using the matrix elements of the ground state $|\Phi_0\rangle$ as listed in Appendix, which can be calculated before the procedure of the determination of the variational parameters.  Once the ground state wavefunction is optimized, these matrix elements can be calculated by using it.
The variational parameters for the exciations, $g^{\rm (A)}_{\delta}, \alpha^{\rm (A)}_{\delta},\beta^{\rm (A)}_{\delta}$ and $\gamma^{\rm (A)}_{\delta}$ are optimized afterwards by using the above matrix elements in the ground state.

\subsubsection{Renormalization Factor and Line Width}
The renormalization factor $Z_i$ for the particle excitation $f_i^{\dagger}$ can be calculated by the inertial product
\begin{eqnarray}
Z_{i}&=&|\langle \Psi_{i}(k)|c_{k,\sigma}^{\dagger}|\Phi_{0}\rangle |^2, \label{eq:Zi}
\end{eqnarray}

The line width is related to the life time of the elementary excitation. Here, the width can be calculated from
\begin{eqnarray}
A_i(k,\omega)&=&\int dt e^{i\omega t} \langle \Phi_{0}|e^{iHt}c_{k\sigma}|\Psi_{i}(k)\rangle \nonumber \\
&&\langle \Psi_{i}(k)|e^{-iHt} c_{k,\sigma}^{\dagger}|\Phi_{0}\rangle, \label{eq:Akw}
\end{eqnarray}
When we assume the ground state energy of $|\Phi_{0}\rangle$, $E_0$, the average over $|\Psi_{i}(k)\rangle$ leads to 
\begin{eqnarray}
A_i(k,\omega)=Z_i \int d\omega e^{i(\omega+E_0) t} \langle \Psi_{i}(k)|e^{-iHt}|\Psi_{i}(k)\rangle, 
\label{eq:Akw2}
\end{eqnarray}
By rewriting $\langle \Psi_{i}(k)|e^{-iHt}|\Psi_{i}(k)\rangle$ as $\langle e^{-iHt}\rangle_i$, it is simplified to 
\begin{eqnarray}
A_i(k,\omega)=Z_i \int d\omega e^{i(\omega+E_0-E_i) t}  e^{-\Delta E_i^2 t^2}, 
\label{eq:Akw3}
\end{eqnarray}
up to the second cumulant, where
$E_i=\langle H \rangle_i$ and $\Delta E_i^2 =\langle H^2 \rangle_i -\langle H \rangle_i^2$.
After the Fourier transform, we obtain a Gaussian
\begin{eqnarray}
A_i(k,\omega)=Z_i \exp[-\frac{(\omega+E_0-E_i)^2}{\Delta E_i^2}], 
\label{eq:Akw4}
\end{eqnarray}
with the width
$\Delta E_i$.

\subsection{Hole Excitation}

The formalism for hole excitations can be obtained by a straightforward extension of the particle excitations.
Note that the optimized excitations are not necessarily the simple particle hole conjugate.  Instead, the quasihole $\tilde{c}^{(h)}_{k\sigma}$, \red{dark} hole $d^{\rm (h)}$ and  Mott-gap hole $d^{\rm (MGh)}$ may be obtained in a procedure similar to the particle excitations but independently of them.
The renormalization factor of a hole is given from 
\begin{eqnarray}
Z_{i}&=&|\langle \Psi_{i}(k)|c_{k,\sigma}|\Phi_{0}\rangle|^2. \label{eq:Zi_hole}
\end{eqnarray}

\section{Discussion}\label{Sec6}
\subsection{Bistability, Attraction and Competing Order}
\begin{figure}[h!]
  \begin{center}
    \includegraphics[width=4cm]{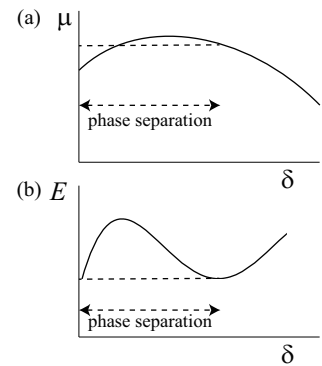}
    \end{center}
\caption{
Schematic illustration for doping concentration dependence of chemical potential and energy. 
Dashed line gives the Maxwell construction for 
the phase separation.
}
\label{Chem-Ene}
\end{figure}
Since quasiparticle (quasihole) and composite fermion (\red{dark} fermion) are both low-energy excitations in both sides of underdoped hole and electron doping regions, they have bistability and the relative stability may change with evolution of doping according as the level shift of $d$ relative to $c$. As we mentioned in the end of Sec.~\ref{Sec3} and in the end of Sec.~\ref{Sec4}, the \red{dark} fermion is expected to be more stable in the smaller doping region and the quasiparticle becomes stable in the higher doping. Accordingly, in the intermediate doping region, the energy to add a hole (electron) becomes relatively high and the chemical potential is expected to have a peak structure schematically shown in Fig.\ref{Chem-Ene}(a) 
which results in the tendency for the phase separation 
indicated by the Maxwell construction. This is also understood from the energy curve in Fig.\ref{Chem-Ene}(b). Such doping dependences have indeed been observed in numerical results\cite{Misawa2014}. The diverging charge compressibility $(d^2E/d\delta^2)^{-1}$ triggers the phase separation~\cite{Furukawa1992} and the resultant phase separation have been extensively studied in other studies as well~\cite{Emery,Cosentini,Capone2006,Aichhorn2007,Khatami2010,Chang2010,Neuscamman2012,Sordi,Yokoyama2013}.
Even in the case of the iron-based superconductors, such tendency has been pointed out in the {\it ab initio} calculation\cite{MisawaFeNcomm}, suggesting a universal underlying mechanism of high $T_{\rm c}$ superconductivity. Recent refined numerical results suggest that the charge inhomogeneity emerges as charge/spin stripes at least in a part of phase separation region~\cite{Himeda,Corboz,Zhao2017,Simon-Foundation,Ido2018}.  
Although the configuration depends on the details (band structure, interaction and its spatial range, doping concentration etc.) the tendency for the electronic inhomogeneity is robust. In real compounds with long-ranged Coulomb interaction, the macroscopic phase separation is of course prohibited and some real space configuration of the charge inhomogeneity including the stripe~\cite{Tranquada1995} and patch structure~\cite{davis} may appear depending on the detailed experimental condition.

Since the energy shown in Fig. \ref{Chem-Ene}(b) has a negative curvature, the total electronic energy is expanded as 
\begin{eqnarray}
E = E_0+a\delta+b\delta^2+\cdots
\label{eq:attraction}
\end{eqnarray} 
in terms of the doping concentration $\delta$ dependence
with $b<0$. This means that the effective carrier interaction is attractive.  
It is reasonable because the phase separation or charge inhomogeneity including the stripe order is driven by the effective attraction of carriers and such attraction simultaneously helps the Cooper pairing and resultant strong-coupling superconductivity,
as is observed numerically as a severe competition of the stripe and charge homogeneous $d$-wave superconducting state\cite{Misawa2014}.
It is the characteristics of the strong-coupling superconductivity, where the retardation effect is small and
the coherence length is small as well, requiring more or less instantaneous attraction of carriers for the Cooper pair formation. This is far different from the conventional BCS superconductivity. 

The bistability is interpreted as the origin of this effective attraction, while it can also be interpreted by the Mottness, where the kinetic energy is lowered in a nonlinear fashion with evolution of doping~\cite{Misawa2014}. In the Mott insulator the kinetic energy gain arising from the itinerancy of electrons is suppressed while its gain increases nonlinearly upon doping by the recovery of the electron coherence. This is equivalent to the switch of the character of the carriers from the \red{dark} particles to quasiparticles with the doping evolution.

Although the carrier consists of single component in the single-band Hubbard model, we find a sort of ``fractionalization" of electrons into the \red{dark} particle and quasiparticle.   However, these two excitations are not a true eigenstate because of their interactions and their hybridization. This can be studied in variational numerical studies in the present formalism. 
In other words, the hybridization introduces the life time of quasiparticle and \red{dark} particle by their mutual transformation, namely the hybridization, generating the two types of hybridization gap, Mott gap and the pseudogap.  Numerical results will be reported elsewhere.

In the literature, several different types of ``fractionalization" have been studied. An example is the slave boson formalism\cite{Kotliar-Ruckenstein,Yamaji,YamajiPRB}. The present \red{dark} fermion has a clear physical picture. In the Mott insulator, the Mott gap fermion $d^{\rm (MG)}$ and the original electron $c$ constitute a well defined two-component system. 
In the atomic limit, the fractionalization is exactly established, where the full Hilbert space of the Hubbard model (namely, interacting fermions) is remarkably and precisely mapped to the Hilbert space of a noninteracting two-component fermion model in the restricted particle number sector.    
The doped system  involves an additional weakly bound exciton as a composite object, not the decomposition of the single electron as in the slave bosons and in principle experimentally detectable.

\subsection{Experimental Challenge}  

To verify the electronic structure unique in the present mechanism, we need a refined experimental tool and analysis beyond the existing probes.  A way to extract the normal and anomalous part of the single-particle electron self-energy separately and examine whether their cancellation occurs in the contribution to the Green's function in the superconducting phase will be a smoking gun for the present mechanism, which has been discussed before~\cite{Sakai2016,Sakai2016_2}. In addition, it is important to figure out the ultrafast charge dynamics to characterize and confirm the present picture of two types of excitons and associated \red{dark} fermion dynamics. In particular, a challenging issue is the detection of the exciton condensation and its dynamics expected in the Mott insulating as well as in the underdoped regions. The resonant inelastic X-ray scattering (RIXS)~\cite{Hill,Abbamonte_RIXS,Hasan,YJKim,Ellis}, Raman scattering~\cite{Chen_Raman,Sakai2013} and momentum-resolved electron energy loss spectroscopy (MEELS)~\cite{abbamonte} are promising tools for this purpose, if the energy and momentum resolutions are sufficient.    
\vspace{5truemm}

\noindent
{\large \bf Acknowledgements}
\vspace{2truemm}
\appendix
We acknowledge useful discussions with Shiro Sakai and Marcello Civelli. 
Numerical data for Fig.1 are provided by Takahiro Misawa by using H$\Phi$ code~\cite{HPhi,HPhi2}.
This work was financially supported by Grant-in-Aids
for Scientific Research (JSPS KAKENHI) (No. 16H06345) from Ministry of Education, Culture,
Sports, Science and Technology (MEXT), Japan. This work was
also supported in part by MEXT as a social and scientific
priority issue (Creation of new functional devices and
high-performance materials to support next-generation
industries (CDMSI)) to be tackled by using post-K computer. We also thank the support by the RIKEN Advanced Institute for Computational Science through
the HPCI System Research project (hp160201,hp170263) supported by MEXT. 

\appendix
\setcounter{secnumdepth}{1}
\section{Mapping between the two-component fermion model and the Hubbard model at half filling in the atomic limit} \label{appendix:AtomicLimit}
We consider the Hubbard $U$ term 
\red{
\begin{eqnarray}
{\cal H}_{U}&=& 
Un_{\uparrow}n_{\downarrow},
\label{eq:atomic_limit}
\end{eqnarray}
}
with $n_{\sigma}=c^{\dagger}_{\sigma}c_{\sigma}$.
We introduce the Mott-gap fermion 
\begin{eqnarray}
\tilde{d}_{\sigma}=c_{\sigma}(1-2n_{-\sigma}),
\label{eq:d}
\end{eqnarray}
together with 
\begin{eqnarray}
\tilde{c}_{\sigma}=c_{\sigma}.
\label{eq:c}
\end{eqnarray}
For the spin $\sigma$ part, Eq.(\ref{eq:atomic_limit}) can be rewritten as
\begin{eqnarray}
{\cal H}_{\rm TCFM}&=& {\cal H}^{(\tilde{c})} +{\cal H}^{(\tilde{d})}+{\cal H}^{(\tilde{c}\tilde{d})}, \label{eq:hamiltonian} \label{eq:HTCFM} \\
{\cal H}^{(\tilde{c})}&=&\epsilon_{\tilde{c}} \tilde{c}_{\sigma}^{\dagger}\tilde{c}_{\sigma}, \\
{\cal H}^{(\tilde{d})}&=& \epsilon_{\tilde{d}}\tilde{d}_{\sigma}^{\dagger}\tilde{d}_{\sigma}, \\
{\cal H}^{(\tilde{c}\tilde{d})}&=& \Lambda (\tilde{c}_{\sigma}^{\dagger}\tilde{d}_{\sigma} + {\rm H.c}),
\end{eqnarray}
\red{with Eq(\ref{cde}).
For the derivation of Eq.(\ref{eq:HTCFM}) see below}. 

Note first that $\tilde{d}$ and $\tilde{c}$ satisfy the exact anticommutation relation in the ground state average,
\begin{eqnarray}
\langle \tilde{c}_{\sigma}\tilde{c}_{\sigma}^{\dagger}+\tilde{c}_{\sigma}^{\dagger}\tilde{c}_{\sigma}\rangle=1, \\
\langle \tilde{d}_{\sigma}\tilde{d}_{\sigma}^{\dagger}+\tilde{d}_{\sigma}^{\dagger}\tilde{d}_{\sigma}\rangle=1, \\
\langle \tilde{c}_{\sigma}\tilde{d}_{\sigma}^{\dagger}+\tilde{d}_{\sigma}^{\dagger}\tilde{c}_{\sigma}\rangle=0,
\label{eq:anticom}
\end{eqnarray}
where
$\langle \cdots \rangle$ is 
\begin{eqnarray}
\langle \cdots \rangle =\frac{\langle \uparrow |\cdots |\uparrow\rangle+\langle \downarrow |\cdots |\downarrow\rangle}{\langle \uparrow |\uparrow\rangle+\langle \downarrow |\downarrow\rangle}.
\label{eq:avearge}
\end{eqnarray}
In this sense, $\tilde{c}$ and $\tilde{d}$ behave as orthogonal fermions as single-particle excitations from the ground state, which is degenerate ensemble of $|\uparrow\rangle$ and $|\downarrow \rangle$.

\red{
By diagonalizing Eq.(\ref{eq:HTCFM}) for the case $\epsilon_{\tilde{c}}=\epsilon_{\tilde{d}}$, one obtains for the spin $\sigma$ part
\begin{eqnarray}
{\cal H}_{\rm DTCFM}&=& \epsilon_a a_{\sigma}^{\dagger}a_{\sigma}+\epsilon_b b_{\sigma}^{\dagger}b_{\sigma},  \label{eq:DHTCFM}\\
\epsilon_b &=& \frac{1}{2}(\epsilon_{\tilde{c}}+\epsilon_{\tilde{d}})+\Lambda, \\
\epsilon_a &=& \frac{1}{2}(\epsilon_{\tilde{c}}+\epsilon_{\tilde{d}})-\Lambda
\end{eqnarray}
with Eqs.(\ref{eq:ab0}) and (\ref{eq:ab}).
With the choice Eq.(\ref{cde}),
$\epsilon_a=0$ and $\epsilon_b=U$ are obtained and the Hubbard gap is reproduced.
}

\red{
Here, we need care about the degeneracy of the ground state at half filling of Eq.(\ref{eq:atomic_limit}),
where the two singly occupied states $|\uparrow\rangle$ and $|\downarrow \rangle$ are degenerate as we take the average as Eq.(\ref{eq:avearge}).  
Therefore the Hilbert space of Eq.(\ref{eq:atomic_limit}) for the single-particle state is given by these two energy eigenstates, where the degenerate energy is 0.  In the Hilbert space of Eq.(\ref{eq:HTCFM}), the mapped state is $a_{\uparrow}|0\rangle$ and $a_{\downarrow}|0\rangle$. Depending on which state is taken for the filling of $\epsilon_a$ level, only one type of $b$ fermion can be filled: If $a_{\sigma}$ fermion is filled, only $b_{-\sigma}$ fermion at the $\epsilon_b$ level can be created if we trace back to the original $c$ operator.  However, we do not need to care about it in the Hilbert space of Eq.(\ref{eq:HTCFM}) when the particle and hole excitations are treated separately.
}

\red{
On the other hand, we need to pay attention to the counting of the state in the following way.
When we substitute Eqs.(\ref{eq:d}) and (\ref{eq:c}) into Eq. (\ref{eq:HTCFM}), one obtains
$ -4\Lambda n_{\sigma}n_{-\sigma}
$
for $\epsilon_{\tilde{c}}=\epsilon_{\tilde{d}}=-\Lambda$.
However, to make the mapping correct, the degeneracy of the half-filled ground state in the Hilbert space of the original single-component fermion in Eq.(\ref{eq:atomic_limit}) as addressed above is required to be taken into account when one calculates physical quantities in the Hilbert space of Eq.(\ref{eq:HTCFM}). Namely, the mapping has to take account of the factor 1/2 coming from the denominator of Eq.(\ref{eq:avearge}). Then Eq.(\ref{eq:HTCFM}) is mapped to 
\begin{eqnarray}
&& U n_{\sigma}n_{-\sigma}
\end{eqnarray}
if $\Lambda=-U/2$, as derived in the spectrum of Eq.(\ref{eq:DHTCFM}).}


In general, one can show
\begin{eqnarray}
 G_{\tilde{c}_{\sigma'},\tilde{c}_{\sigma'}^{\dagger}} (\omega ) &=& \frac{1}{\omega-\epsilon_{\tilde{c}}-\frac{\Lambda^2}{\omega-\epsilon_{\tilde{d}}}},
\nonumber \\
G_{\tilde{c}_{\sigma'},\tilde{d}_{\sigma'}^{\dagger}}(\omega) &=& G_{\tilde{d}_{\sigma'},\tilde{c}_{\sigma'}^{\dagger}}(\omega) 
=\frac{-\Lambda}{(\omega-\epsilon_{\tilde{c}})(\omega-\epsilon_{\tilde{d}})-\Lambda^2}\nonumber \\
G_{\tilde{d}_{\sigma'},\tilde{d}_{\sigma'}^{\dagger}}(\omega)&=& \frac{1}{\omega-\epsilon_{\tilde{d}}-\frac{\Lambda^2}{\omega-\epsilon_{\tilde{c}}}},
\label{eq:Gtildectilded}
\end{eqnarray}

\red{
Then from Eq.(\ref{cde}), we obtain
\begin{eqnarray}
 G_{\tilde{c}_{\sigma'},\tilde{c}_{\sigma'}^{\dagger}} (\omega ) &=& \frac{1}{\omega-\frac{U}{2}-\frac{\frac{U^2}{4}}{\omega-\frac{U}{2}}} \\
 &=&\frac{1}{2}\left[ \frac{1}{\omega}+\frac{1}{\omega-U}\right]
\label{eq:G_ctilde}
\end{eqnarray}
This is equivalent to the Green's function of Eq.(\ref{eq:atomic_limit}) when we shift the constant energy $\omega-U/2 \rightarrow \omega$.
The constant energy shift may be interpreted as that to adjust the chemical potential. 
The self-energy has the correct form as well:
\begin{eqnarray}
{\Sigma}(\omega)&=&\frac{\frac{U^2}{4}}{\omega-\frac{U}{2}}.
\label{eq:Sigma}
\end{eqnarray}
}

In this way, exact correspondence is established between the two-component fermion model (\ref{eq:hamiltonian}) and the Hubbard model for the half-filled ground state as well as for single-particle excitations from it. Namely, the full Hilbert space of the Hubbard model in the atomic limit is equivalent to the two-component fermion model in the particle number sector from 1 to 3.
\section{Matrix Elements for Construction of Wavefunction}
In practical calculations, we need to calculate 

\begin{eqnarray}
\Gamma_1(j;n) &=& \langle{c}_{j,\sigma}{c}_{n,\sigma}^{\dagger} \rangle,   \label{eq:Gamma1}
\end{eqnarray}
\begin{multline}
\Gamma_2(j,\delta_1,\sigma';\sigma;n,\delta_2,\sigma'')
 = \langle{n}_{j+\delta_1, \sigma'}{c}_{j,\sigma}{c}_{n,\sigma}^{\dagger}{n}_{n+\delta_2,\sigma''}\rangle,    
\end{multline}
\begin{eqnarray}
\Gamma_3(j;\sigma;n,\delta_2,\sigma') &=& \langle{c}_{j,\sigma}{c}_{n,\sigma}^{\dagger}{n}_{n+\delta_2,\sigma'}  \rangle,  \\
\Gamma_4(j,\delta_1,\sigma';\sigma;n) &=& \langle{n}_{j+\delta_1,\sigma'}{c}_{j, \sigma}{c}_{n,\sigma}^{\dagger}
  \rangle,  \\
\Gamma_5(j;n) &=& \langle{c}_{j,\sigma}^{\dagger}{c}_{n,\sigma} \rangle,   
\end{eqnarray}
\begin{multline}
\Gamma_6(j,\delta_1,\sigma';\sigma;n,\delta_2,\sigma'') = \langle{n}_{j+\delta_1, \sigma'}{c}_{j,\sigma}^{\dagger}{c}_{n,\sigma}{n}_{n+\delta_2,\sigma''}\rangle,  
\end{multline}
\begin{eqnarray}
\Gamma_7(j;\sigma;n,\delta_2,\sigma') &=& \langle{c}_{j,\sigma}^{\dagger}{c}_{n,\sigma}{n}_{n+\delta_2,\sigma'}  \rangle,   \\
\Gamma_8(j,\delta_1,\sigma';\sigma;n) &=& \langle{n}_{j+\delta_1,\sigma'}{c}_{j, \sigma}^{\dagger}{c}_{n,\sigma}
  \rangle, 
\end{eqnarray}
\begin{multline}
\Gamma_9(j,\delta_1;\sigma;n,\delta_2) \\
 = \langle{n}_{j+\delta_1, \sigma}{n}_{j+\delta_1, -\sigma}{c}_{j,\sigma}{c}_{n,\sigma}^{\dagger} \rangle, 
\label{eq:Gamma9} 
\end{multline}
\begin{multline}
\Gamma_{10}(j,\delta_1;\sigma;n,\delta_2,\sigma') \\
 = \langle{n}_{j+\delta_1, \sigma}{n}_{j+\delta_1, -\sigma}{c}_{j,\sigma}{c}_{n,\sigma}^{\dagger} {n}_{n+\delta_2,\sigma'})\rangle, \label{eq:Gamma10} 
\end{multline}
\begin{multline}
\Gamma_{11}(j,\delta_1;n,\delta_2) \\
 = \langle{n}_{j+\delta_1, \sigma}{n}_{j+\delta_1, -\sigma}{c}_{j,\sigma}{c}_{n,\sigma}^{\dagger} 
{n}_{n+\delta_2,-\sigma}{n}_{n+\delta_2,\sigma}\rangle, \label{eq:Gamma11} 
\end{multline}
\begin{multline}
\Gamma_{12}(j,\delta_1;\sigma;n,\delta_2) \\
 = \langle{n}_{j+\delta_1, \sigma}{n}_{j+\delta_1, -\sigma}{c}_{j, \sigma}^{\dagger}{c}_{n,\sigma} \rangle, 
\label{eq:Gamma12} 
\end{multline}
\begin{multline}
\Gamma_{13}(j,\delta_1;\sigma;n,\delta_2,\sigma') \\
 = \langle{n}_{j+\delta_1, \sigma}{n}_{j+\delta_1, -\sigma}{c}_{j, \sigma}^{\dagger}{c}_{n,\sigma} {n}_{n+\delta_2,\sigma'})\rangle, \label{eq:Gamma13} 
\end{multline}
\begin{multline}
\Gamma_{14}(j,\delta_1;n,\delta_2) \\
 = \langle{n}_{j+\delta_1, \sigma}{n}_{j+\delta_1, -\sigma}{c}_{j, \sigma}^{\dagger}{c}_{n,\sigma} 
{n}_{n+\delta_2,-\sigma}{n}_{n+\delta_2,\sigma}\rangle, \label{eq:Gamma14} 
\end{multline}
\begin{eqnarray}
\Xi_1(j;n) &=& \langle{c}_{j,\sigma}H{c}_{n,\sigma}^{\dagger} \rangle,   \label{eq:Xi1}
\end{eqnarray}
\begin{multline}
\Xi_2(j,\delta_1,\sigma';\sigma;n,\delta_2,\sigma'')
 = \langle{n}_{j+\delta_1, \sigma'}{c}_{j,\sigma}H{c}_{n,\sigma}^{\dagger}{n}_{n+\delta_2,\sigma''}\rangle,    
\end{multline}
\begin{eqnarray}
\Xi_3(j;\sigma;n,\delta_2,\sigma') &=& \langle{c}_{j,\sigma}H{c}_{n,\sigma}^{\dagger}{n}_{n+\delta_2,\sigma'}  \rangle,  \\
\Xi_4(j,\delta_1,\sigma';\sigma;n) &=& \langle{n}_{j+\delta_1,\sigma'}{c}_{j, \sigma}H{c}_{n,\sigma}^{\dagger}
  \rangle,  \\
\Xi_5(j;n) &=& \langle{c}_{j,\sigma}^{\dagger}H{c}_{n,\sigma} \rangle,   
\end{eqnarray}
\begin{multline}
\Xi_6(j,\delta_1,\sigma';\sigma;n,\delta_2,\sigma'') = \langle{n}_{j+\delta_1, \sigma'}{c}_{j,\sigma}^{\dagger}H{c}_{n,\sigma}{n}_{n+\delta_2,\sigma''}\rangle,  
\end{multline}
\begin{eqnarray}
\Xi_7(j;\sigma;n,\delta_2,\sigma') &=& \langle{c}_{j,\sigma}^{\dagger}H{c}_{n,\sigma}{n}_{n+\delta_2,\sigma'}  \rangle,   \\
\Xi_8(j,\delta_1,\sigma';\sigma;n) &=& \langle{n}_{j+\delta_1,\sigma'}{c}_{j, \sigma}^{\dagger}H{c}_{n,\sigma}
  \rangle, 
\end{eqnarray}
\begin{multline}
\Xi_9(j,\delta_1;\sigma;n,\delta_2) \\
 = \langle{n}_{j+\delta_1, \sigma}{n}_{j+\delta_1, -\sigma}{c}_{j,\sigma}H{c}_{n,\sigma}^{\dagger} \rangle, 
\label{eq:Xi9} 
\end{multline}
\begin{multline}
\Xi_{10}(j,\delta_1;\sigma;n,\delta_2,\sigma') \\
 = \langle{n}_{j+\delta_1, \sigma}{n}_{j+\delta_1, -\sigma}{c}_{j,\sigma}H{c}_{n,\sigma}^{\dagger} {n}_{n+\delta_2,\sigma'})\rangle, \label{eq:Xi10} 
\end{multline}
\begin{multline}
\Xi_{11}(j,\delta_1;n,\delta_2) \\
 = \langle{n}_{j+\delta_1, \sigma}{n}_{j+\delta_1, -\sigma}{c}_{j,\sigma}H{c}_{n,\sigma}^{\dagger} 
{n}_{n+\delta_2,-\sigma}{n}_{n+\delta_2,\sigma}\rangle, \label{eq:Xi11} 
\end{multline}
\begin{multline}
\Xi_{12}(j,\delta_1;\sigma;n,\delta_2) \\
 = \langle{n}_{j+\delta_1, \sigma}{n}_{j+\delta_1, -\sigma}{c}_{j, \sigma}^{\dagger}H{c}_{n,\sigma} \rangle, 
\label{eq:Xi12} 
\end{multline}
\begin{multline}
\Xi_{13}(j,\delta_1;\sigma;n,\delta_2,\sigma') \\
 = \langle{n}_{j+\delta_1, \sigma}{n}_{j+\delta_1, -\sigma}{c}_{j, \sigma}^{\dagger}H{c}_{n,\sigma} {n}_{n+\delta_2,\sigma'})\rangle, \label{eq:Xi13} 
\end{multline}
\begin{multline}
\Xi_{14}(j,\delta_1;n,\delta_2) \\
 = \langle{n}_{j+\delta_1, \sigma}{n}_{j+\delta_1, -\sigma}{c}_{j, \sigma}^{\dagger}H{c}_{n,\sigma} 
{n}_{n+\delta_2,-\sigma}{n}_{n+\delta_2,\sigma}\rangle, \label{eq:Xi14} 
\end{multline}
where the average of an operator $A$, $\langle A \rangle$ is defined as 
\begin{eqnarray}
\langle A \rangle=\frac{\langle \Phi_0 |A|\Phi_0 \rangle}{\langle \Phi_0 |\Phi_0 \rangle}.
\label{eq:ave}
\end{eqnarray}
We note the symmetries such as
\begin{eqnarray}
\Theta_1(j;n)&=& \Theta_1(n;j)^*, \\
\Theta_2(j,\delta_1;n,\delta_2)&=& \Theta_2(n,\delta_2,j,\delta_1)^*,  \\
\Theta_3(j;n,\delta_2)&=&\Theta_4(n,\delta_2;j)^*,  \\
\Theta_5(j;n)&=& \Theta_5(n;j)^*, \\
\Theta_6(j,\delta_1;n,\delta_2)&=& \Theta_6(n,\delta_2,j,\delta_1)^*,  \\
\Theta_7(j;n,\delta_2)&=&\Theta_8(n,\delta_2;j)^*,  
\end{eqnarray}
\begin{multline}
\Theta_9(j,\delta_1;n,\delta_2,\alpha_{\delta_1},\alpha_{\delta_2},\beta_{\delta_1},\beta_{\delta_2}^{\rm (h)}) \\
=\Theta_9(n,\delta_2;j,\delta_1,\alpha_{\delta_2},\alpha_{\delta_1},\beta_{\delta_2},\beta_{\delta_1})^*, 
\end{multline}
\begin{multline}
\Theta_{10}(j,\delta_1;n,\delta_2,\alpha_{\delta_1}^{\rm (h)},\alpha_{\delta_2}^{\rm (h)},\beta_{\delta_1}^{\rm (h)},\beta_{\delta_2}^{\rm (h)}) \\
=\Theta_{10}(n,\delta_2;j,\delta_1,\alpha_{\delta_2}^{\rm (h)},\alpha_{\delta_1}^{\rm (h)},\beta_{\delta_2}^{\rm (h)},\beta_{\delta_1}^{\rm (h)})^*,  
\label{eq:Thetasymm}
\end{multline}
where $\Theta$ is either $\Gamma$ or $\Xi$.

The Fourier transform is defined  for both $\Theta=\Gamma$ and $\Theta=\Xi$ by
\begin{eqnarray}
\Theta_m(k) &=& \frac{1}{\sqrt{N}}\sum_j \exp[ik(r_j-r_n)]\Theta_m(j,n), \nonumber \\
&& \ \ \ \ \ {\rm for \ } m= 1, 5, \\
\Theta_m(k,\delta_{i}) &=& \frac{1}{\sqrt{N}}\sum_j \exp[ik(r_j-r_n)]\Theta_m(j;n,\delta_{i}), \nonumber \\
&& \ \ \ \ \ {\rm for \ } m= 3, 7, \\
\Theta_m(k,\delta_{i}) &=& \frac{1}{\sqrt{N}}\sum_j \exp[ik(r_j-r_n)]\Theta_m(j,\delta_{i};n),  \nonumber \\
&& \ \ \ \ \ {\rm for \ } m= 4, 8,\\
\Theta_m(k,\delta_{i},\delta_{l}) &=&\frac{1}{\sqrt{N}}\sum_j \exp[ik(r_j-r_n)]\Theta_m(j,\delta_{i};n,\delta_l) \nonumber \\
&& \ \ \ \ \ {\rm for \ } m= 2, 6, 
\end{eqnarray}
\begin{multline}
\Theta_m(k,\delta_{i},\delta_{l};\alpha_{\delta_{i}},\alpha_{\delta_{l}},\beta_{\delta_{i}},\beta_{\delta_{l}}) \\
=\frac{1}{\sqrt{N}}\sum_j \exp[ik(r_j-r_n)]\Theta_m(j,\delta_{i};n,\delta_l;\alpha_{\delta_{i}},\alpha_{\delta_{l}},\beta_{\delta_{i}},\beta_{\delta_{l}})  \\  {\rm for \ } m=  9, 10, 
\label{eq:XG}
\end{multline}
We used the translational invariance: $\Xi_{m}(j,n)$ and $\Gamma_{m}(j,n)$ depend only on $r_j-r_n$.


\begin{thebibliography}{99}
\bibitem{Yasuoka}
H. Yasuoka, T. Imai, and T. Shimizu, Strong Correlation and
Superconductivity: Proceedings of the IBM Japan International
Symposium, Mt. Fuji, Japan, pp.21-25 May, 1989 (Springer Berlin
Heidelberg) , 254 (1989).
\bibitem{Alloul}
H. Alloul, T. Ohno, and P. Mendels, Phys. Rev. Lett. {\bf 63}, 1700
(1989).
\bibitem{RMP}
M. Imada, A. Fujimori, and Y. Tokura, Rev. Mod. Phys. {\bf 70}, 1039
(1998).
\bibitem{Tranquada1995} J. M.Tranquada, B. J. Sternlieb, J. D.Axe, Y. Nakamura, and S. Uchida: Nature {\bf 375}, 561 (1995).
\bibitem{Tranquada1997} J. M. Tranquada, J. D. Axe, N. Ichikawa, A. R. Moodenbaugh, Y. Nakamura, and S. Uchida: Phys. Rev. Lett. {\bf 78}, 338 (1997).
\bibitem{Yamada1998} K. Yamada, C. H. Lee, K. Kurahashi, J. Wada, S. Wakimoto, S. Ueki,
H. Kimura, Y. Endoh, S. Hosoya, G. Shirane, R. J. Birgeneau,
M. Greven, M. A. Kastner, and Y. J. Kim: Phys. Rev. B {\bf 57}, 6165 (1998).
\bibitem{Fink2011} J. Fink, V. Soltwisch, J. Geck, E. Schierle, E.Weschke, and B. B\"{u}chner, Phys. Rev. B {\bf 83}, 092503 (2011).
\bibitem{Ghiringhelli}  G. Ghiringhelli, M. Le Tacon, M. Minola, S. Blanco-Canosa, C. Mazzoli,
N. B. Brookes, G. M. De Luca, A. Frano, D. G. Hawthorn, F. He,
T. Loew, M. M. Sala, D. C. Peets, M. Salluzzo, E. Schierle, R. Sutarto,
G. A. Sawatzky, E. Weschke, B. Keimer, and L. Braicovich: Science
{\bf 337}, 821 (2012) .
\bibitem{Comin} R. Comin, R. Sutarto, E. H. da Silva Neto, L. Chauviere, R. Liang,
W. N. Hardy, D. A. Bonn, F. He, G. A. Sawatzky, and A. Damascelli:
Science {\bf 347}, 1335 (2015).
\bibitem{Peng}  Y. Y. Peng, M. Salluzzo, X. Sun, A. Ponti, D. Betto, A. M. Ferretti,
F. Fumagalli, K. Kummer, M. Le Tacon, X. J. Zhou, N. B. Brookes,
L. Braicovich, and G. Ghiringhelli: Phys. Rev. B {\bf 94}, 184511 (2016).
\bibitem{SatoNematic} Y. Sato, S. Kasahara, H. Murayama, Y. Kasahara, E.-G. Moon, T. Nishizaki, T. Loew, J. Porras, B. Keimer, T. Shibauchi, Y. Matsuda, Nat. Phys. {\bf 13}, 1074 (2017).

\bibitem{davis} 
S.H. Pan {\it et al.}, Nature {\bf 413} 282 (2001)
\bibitem{Maier} T. A. Maier, T. Pruschke, and M. Jarrell, Phys. Rev. B {\bf 66}, 075102
(2002).
\bibitem{Senechal} D. \'{S}e\'{n}echal and A.-M. S. Tremblay, Phys. Rev. Lett. {\bf 92}, 126401
(2004).
\bibitem{Civelli2005} M. Civelli, M. Capone, S. S. Kancharla, O. Parcollet, and G. Kotliar, Phys. Rev. Lett. 95, 106402 (2005).
\bibitem{Berthod} C. Berthod, T. Giamarchi, S. Biermann, and A. Georges, Phys.
Rev. Lett. 97, 136401 (2006).
\bibitem{Stanescu} T. D. Stanescu and G. Kotliar, Phys. Rev. B {\bf 74}, 125110 (2006).
\bibitem{Konik} R. M. Konik, T. M. Rice, and A. M. Tsvelik, Phys. Rev. Lett. {\bf 96},
086407 (2006).
\bibitem{YRZ} K.-Y. Yang, T. M. Rice, and F.-C. Zhang, Phys. Rev. B {\bf 73},
174501 (2006).
\bibitem{HauleKotliar} K. Haule and G. Kotliar, Phys. Rev. B {\bf 76}, 104509 (2007).
\bibitem{Norman}
Phys. Rev. B {\bf 76}, 174501 (2007).
\bibitem{Civelli2009}
M. Civelli, Phys. Rev. B, Phys. Rev. B {\bf 79}, 195113 (2009).
\bibitem{Sakai2009}
S. Sakai, Y. Motome and M. Imada,
Phys. Rev. Lett. {\bf 102}, 056404 (2009).
\bibitem{Civelli2009PRL}
M. Civelli, Phys. Rev. Lett. 103, 136402 (2009).
\bibitem{Sakai2010}
S. Sakai, Y. Motome and M. Imada,
Phys. Rev. B {\bf 82}, 134505 (2010).
\bibitem{Gull}E. Gull and A. J. Millis, Phys. Rev. B 91, 085116 (2015).
\bibitem{Sakai2016}
S. Sakai, M. Civelli, and M. Imada, Phys. Rev. Lett. {\bf 116}, 057003 (2016).
\bibitem{Sakai2016_2}
S. Sakai, M. Civelli, and M. Imada, Phys. Rev. B {\bf 94}, 115130 (2016).
\bibitem{Weinberg} S. Weinberg {\it Quantum Theory of Fields}, I,II,III (Cambridge University Press, Cambridge, 1995).
\bibitem{Nambu-Jona_Lassinio} Y. Nambu Jona-Lassinio, Phys, Rev. (1963).
\bibitem{Zhu-Zhu} L. Zhu and J.-X. Zhu, Phys. Rev. B {\bf 87}, 085120 (2013).
\bibitem{kotliarCDMFT} G. Kotliar, S. Y. Savrasov, G. PLalsson, and G. Biroli, Phys. Rev.
Lett. {\bf 87}, 186401 (2001).
\bibitem{Kyung} B. Kyung, S. S. Kancharla, D. SLenLechal, A.-M. S. Tremblay, M. Civelli, and G. Kotliar, Phys. Rev. B 73, 165114 (2006).
\bibitem{ZhangImada} Y. Z. Zhang and M. Imada, Phys. Rev. B {\bf 76} 045108 (2007).
\bibitem{Kusunose} H. Kusunose, J. Phys. Soc. Jpn. {\bf 75}, 054713 (2006).
\bibitem{Liebsch} A. Liebsch and N. H. Tong, Phys. Rev. B {\bf 80} 165126 (2009).
\bibitem{Kotliar-Ruckenstein} G. Kotliar and A. E. Ruckenstein, Phys. Rev. Lett {\bf 57}, 1362 (1986).
\bibitem{Yamaji} Y. Yamaji, and M Imada, Phys. Rev. Lett. {\bf 106}, 016404 (2011).
\bibitem{YamajiPRB}
Y. Yamaji, and M Imada, Phys. Rev. B {\bf 83}, 214522 (2011).
\bibitem{Sordi} G. Sordi, P. Semon, K. Haule, and A. M. S. Tremblay, Phys. Rev. Lett. {\bf 108} 216401 (2012).
\bibitem{Gull2013} E. Gull, O. Parcollet, and A.J. Millis, Phys. Rev. Lett. {\bf 110} 216405 (2013).
\bibitem{Sadovskii2005} M. V. Sadovskii, I. A. Nekrasov, E.Z. Kuchinskii, T. Pruschke and V. I. Anisimov, Phys. Rev. B {\bf 72} 155105 (2005).
\bibitem{Daou} R. Daou et al. Nature {\bf 463}, 519 (2010).
\bibitem{Lawrer} M. J. Lawrer {\it et al.}, Nature {\bf 466}, 347 (2010).
\bibitem{Boeuf} D. LeBoeuf et al. Nat. Phys. {\bf 9} 79 (2013).
\bibitem{flux} S. Chakravarty, R. B. Laughlin, D. K. Morr, C. Nayak, Phys. Rev. B {\bf 63}, 094503 (2001).
\bibitem{MaierPoilblancScalapino} T. A. Maier, D. Poilblanc and D. J. Scalapino, Phys. Rev. Lett. {\bf 100}, 237001 (2008). 

\bibitem{Wrobel-Eder} P. Wrobel and R. Eder, Phys. Rev. B {\bf 66}, 035111 (2002).

\bibitem{Rademaker_Zaanen_exciton} L. Rademaker, J. van den Brink, J. Zaanen, and H. Hilgenkamp, Phys. Rev. B {\bf 88}, 235127 (2013).

\bibitem{Imada1986} M. Imada and M. Takahashi, J. Phys. Soc. Jpn. 55, 3354 (1986).
\bibitem{Sugiura} S. Sugiura and A. Shimizu, Phys. Rev. Lett. 108, 240401 (2012).
\bibitem{HPhi} http://ma.cms-initiative.jp/en/application-list/hphi
\bibitem{HPhi2} M. Kawamura, K. Yoshimi, T. Misawa, Y. Yamaji, S. Todo, N. Kawashima 
Comp. Phys. Comm. 217, 180 (2017). 
\bibitem{MisawaYamaji} T. Misawa and Y. Yamaji, J. Phys. Soc. Jpn. {\bf 87}, 023707 (2018).
\bibitem{takagi} H. Takagi, S. Uchida and Y. Tokura, Phys. Rev. Lett. {\bf 62}, 1197 (1989). 
\bibitem{Tahara-Imada}
D. Tahara and M. Imada, J. Phys Soc. Jpn. {\bf 77 },   114701 (2008).
\bibitem{mVMC} https://github.com/issp-center-dev/mVMC
\bibitem{mVMC2} T. Misawa, S. Morita, K. Yoshimi, M. Kawamura, Y. Motoyama, K. Ido, T. Ohgoe, M. Imada, T. Kato, arXiv:1711.11418.
\bibitem{Zhao2017} H.-H. Zhao, K. Ido, S. Morita and M. Imada, Phys. Rev. B. {\bf 96}, 085103 (2017). 
\bibitem{Nomura2017} Y. Nomura, A. S. Darmawan, Y. Yamaji, and M. Imada, 
Phys. Rev. B.{\bf 96}, 205152 (2017).
\bibitem{Imada} M. Imada, S. Sakai, Y. Yamaji and Y. Motome, 
J. Phys. Conf. Ser. {\bf 449}, 012005 (2013).
\bibitem{Misawa2014} T. Misawa and M. Imada: Phys. Rev. B 90 (2014) 115137.

\bibitem{Furukawa1992} N. Furukawa and M. Imada, J. Phys. Soc. Jpn. {\bf 61}, 3331(1992).

\bibitem{Emery} V. J. Emery, S. A. Kivelson, and H. Q. Lin: Phys. Rev. Lett. {\bf 64}, 475 (1990).

\bibitem{Cosentini} A. C. Cosentini, M. Capone, L. Guidoni, and G. B. Bachelet: Phys. Rev. B {\bf 58} (1998) R14685.

\bibitem{Aichhorn2007} M. Aichhorn, E. Arrigoni, M. Potthoff and W. Hanke, Phys. Rev. B {\bf 76}, 224509 (2007).

\bibitem{Khatami2010} E. Khatami, K. Mikelsons,
D. Galanakis, A. Macridin, J. Moreno, R. T. Scalettar, and M. Jarrell, Phys. Rev. B {\bf 81}, 201101(R) (2010).

\bibitem{Capone2006} M. Capone and G. Kotliar, Phys. Rev. B {\bf 74}, 054513 (2006).

\bibitem{Chang2010} C.-C. Chang and S. Zhang, Phys. Rev. Lett. {\bf 104}, 116402 (2010).

\bibitem{Neuscamman2012} E. Neuscamman, C. J. Umrigar, and Garnet Kin-Lic Chan, Phys.
Rev. B {\bf 85}, 045103 (2012).
\bibitem{Yokoyama2013} H. Yokoyama, M. Ogata, Y. Tanaka, K. Kobayashi and H. Tsuchiura, J. Phys. Soc. Jpn. {\bf 82}, 014707 (2013).

\bibitem{MisawaFeNcomm} T. Misawa and M. Imada, Nat. Commun. {\bf 5}, 5738 (2014).
\bibitem{Himeda} A. Himeda, T. Kato, and M. Ogata, Phys. Rev. Lett. {\bf 88}, 117001 (2002).
\bibitem{Corboz} P. Corboz, T. M. Rice, and M. Troyer: Phys. Rev. Lett. {\bf 113},
046402 (2014).
\bibitem{Simon-Foundation} B.-X. Zheng, C.-M. Chung, P. Corboz, G. Ehlers, M.-P. Qin,
R. M. Noack, H. Shi, S. R. White, S. Zhang, and G. K.-L. Chan,
Science {\bf 358}, 1155 (2017).
\bibitem{Ido2018} K. Ido, T. Ohgoe, and M. Imada: Phys. Rev. B {\bf 97}, 045138 (2018).
  
\bibitem{Hill} J. P. Hill, C. C. Kao, W. A. L. Caliebe, M. Matsubara, A. Kotani, J. L. Peng, and R. L. Greene, 
Phys. Rev. Lett. {\bf 80}, 4967 (1998).

\bibitem{Abbamonte_RIXS} P. Abbamonte, C. A. Burns, E. D. Isaacs, P. M. Platzman, PM, L. L. Miller, S.  Cheong, M. V. Klein,  Phys. Rev. Lett. {\bf 83}, 860 (1999).

\bibitem{Hasan}  M. Z. Hasan, E. D. Isaacs, Z. X. Shen, L. L.  Miller, K. Tsutsui, T. Tohyama, S. Maekawa, S, Science {\bf 288}, 1811 (2000).

\bibitem{YJKim} Y. J. Kim, J. P. Hill, C. A. Burns, S.  Wakimoto, R. J. Birgeneau, D. Casa, T. Gog, C. T.  Venkataraman, Phys Rev. Lett. {\bf 89}, 177003 (2002).

\bibitem{Ellis} D. S. Ellis, J. P. Hill, S. Wakimoto, R. J.  Birgeneau, D. Casa, 
T. Gog, Y. J.  Kim,  Phys. Rev. B {\bf 77}, 060501 (2008).

\bibitem{Chen_Raman} X. K. Chen, J. G. Naeini, K. C.  Hewitt, J. C. Irwin, R. Liang, and W. N. Hardy, Phys. Rev. B {\bf 56}, R513 (1997).

\bibitem{Sakai2013} S. Sakai, S. Blanc, M. Civelli, Y. Gallais, M. Cazayous, M.-A. Measson, J. Wen, Z. Xu, G. Gu, G. Sangiovanni, Y. Motome, K. Held, A. Sacuto, A. Georges, and M. Imada,
Phys. Rev. B {\bf 87}, 195144 (2013).

\bibitem{abbamonte} S. Vig, A. Kogar, M. Mitrano, A. A. Husain, V. Mishra, M. S. Rak, L. Venema, P. D. Johnson, G. D. Gu, E. Fradkin, M. R. Norman, P. Abbamonte, SciPost Phys. {\bf 3}, 026 (2017).



\end{thebibliography}
\end{document}